\documentclass{article}

\usepackage{arxiv}

\usepackage[utf8]{inputenc} 
\usepackage[T1]{fontenc}    
\usepackage{hyperref}       
\hypersetup{colorlinks=true, linkcolor=gnblue4, breaklinks=true, urlcolor=gnblue4, citecolor=gnblue4}

\usepackage{url}            
\usepackage{booktabs}       
\usepackage{amsfonts}       
\usepackage{nicefrac}       
\usepackage{microtype}      
\usepackage{lipsum}		
\usepackage{graphicx}
\usepackage[dvipsnames]{xcolor}
\usepackage{natbib}
\usepackage{doi}
\usepackage{titletoc}

\usepackage{bbm}
\usepackage{mathbbol}
\usepackage{imakeidx}

\usepackage{amsmath, amssymb, amsthm}
\usepackage{accents}
\usepackage{upgreek}
\newlength{\dhatheight}

\usepackage{mathtools}

\DeclareMathAlphabet{\mymathbb}{U}{BOONDOX-ds}{m}{n}
\usepackage{yhmath}
\usepackage{mathrsfs}
\usepackage{mathdots}
\usepackage{subfig}
\usepackage[makeroom]{cancel}

\usepackage{tikz}
\usetikzlibrary{matrix, backgrounds, calc}
\usetikzlibrary{graphs, quotes}
\usepackage{geometry}
\usepackage{titlesec}

\newtheorem{definition}{Definition}
\newtheorem{remark}{Remark}
\newtheorem{assumption}{Assumption}
\newtheorem{theorem}{Theorem}
\newtheorem{lemma}{Lemma}
\newtheorem{corollary}[theorem]{Corollary}
\newtheorem{proposition}{Proposition}

\MakeRobust{\eqref}

\definecolor{gnred1}{RGB}{71,0,0} 
\definecolor{gnred2}{RGB}{117,0,0} 
\definecolor{gnred3}{RGB}{164,0,0} 
\definecolor{gnred4}{RGB}{211,0,0} 
\definecolor{gnred5}{RGB}{255,0,0} 
\definecolor{gnred6}{RGB}{255,42,34} 
\definecolor{gnred7}{RGB}{255,91,89} 

\definecolor{gnblue1}{RGB}{0,36,71}   
\definecolor{gnblue2}{RGB}{0,60,118}  
\definecolor{gnblue3}{RGB}{0,85,164}  
\definecolor{gnblue4}{RGB}{0,108,212} 
\definecolor{gnblue5}{RGB}{0,133,255}  
\definecolor{gnblue6}{RGB}{35,156,255} 
\definecolor{gnblue7}{RGB}{88,177,255} 

\definecolor{gnbrown1}{RGB}{71,27,0}  
\definecolor{gnbrown2}{RGB}{117,45,0} 
\definecolor{gnbrown3}{RGB}{164,62,0} 
\definecolor{gnbrown4}{RGB}{211,80,0} 
\definecolor{gnbrown5}{RGB}{255,97,0} 
\definecolor{gnbrown6}{RGB}{255,127,26} 
\definecolor{gnbrown7}{RGB}{255,155,86} 

\newcommand{\En}{\mathrm{E}}

\newcommand{\W}{\operatorname{W}}

\newcommand{\1}{\mathbf{1}}

\newcommand{\real}{\mathbb{R}}

\renewcommand{\real}{\mathbb{R}}

\newcommand{\diag}{\mathrm{diag}}

\usepackage{amsmath,amsfonts,bm}

\def\eqref#1{equation~\ref{#1}}

\def\1{\bm{1}}

\def\ve{{\bm{e}}}

\def\mI{{\bm{I}}}

\DeclareMathAlphabet{\mathsfit}{\encodingdefault}{\sfdefault}{m}{sl}
\SetMathAlphabet{\mathsfit}{bold}{\encodingdefault}{\sfdefault}{bx}{n}

\DeclareSymbolFont{bbold}{U}{bbold}{m}{n}
\DeclareSymbolFontAlphabet{\mathbbold}{bbold}

\newcommand{\vect}[1]{\mathbbold{#1}}
\newcommand{\vectorones}[1][]{\vect{1}_{#1}}
\newcommand{\vectorzeros}[1][]{\vect{0}_{#1}}

\title{Memory flows: geometry and dynamics of sequential retrieval in input-driven Hopfield networks}

\renewcommand{\headeright}{}
\renewcommand{\undertitle}{}
\renewcommand{\shorttitle}{Memory flows: geometry and dynamics of sequential retrieval in input-driven Hopfield networks}

\hypersetup{
pdftitle={Memory flows: geometry and dynamics of sequential retrieval in input-driven Hopfield networks},
pdfauthor={Simone Betteti, Matteo Martin, Morten G. Pedersen, Sandro Zampieri, Giacomo Baggio},
pdfkeywords={Hopfield, Sequential Retrieval, GSPT},
}

\author{ \href{https://orcid.org/0009-0000-3444-0838}{\includegraphics[scale=0.06]{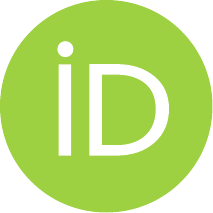}\hspace{1mm}Simone Betteti} \\
	Robust and Intelligent Autonomous Systems (RIAS) lab\\
	The Italian Institute of Artificial Intelligence (AI4I)\\
	\texttt{simone.betteti[at]ai4i.it} 
	\AND
	\href{https://orcid.org/0009-0002-5057-0469}{\includegraphics[scale=0.06]{orcid.pdf}\hspace{1mm}Matteo Martin} \\
	Department of Information Engineering\\
	Università degli Studi di Padova\\
	Padova, 35131, IT \\
	\texttt{matteo.martin[at]studenti.unipd.it} \\
	\And
	\href{https://orcid.org/0000-0002-2639-2394}{\includegraphics[scale=0.06]{orcid.pdf}\hspace{1mm}Morten Gram Pedersen} \\
	Department of Information Engineering\\
	Padua Neuroscience Center (PNC)\\
	Università degli Studi di Padova\\
	Padova, 35131, IT \\
	\texttt{mortengram.pedersen[at]unipd.it}\\
	\AND
	\href{https://orcid.org/0000-0001-8926-1888}{\includegraphics[scale=0.06]{orcid.pdf}\hspace{1mm}Sandro Zampieri} \\
	Department of Information Engineering\\
	Università degli Studi di Padova\\
	Padova, 35131, IT \\
	\texttt{zampi[at]dei.unipd.it}\\
	\And
    \href{https://orcid.org/0000-0002-9439-296X}{\includegraphics[scale=0.06]{orcid.pdf}\hspace{1mm}Giacomo Baggio} \\
	Department of Information Engineering\\
	Università degli Studi di Padova\\
	Padova, 35131, IT \\
	\texttt{baggio[at]dei.unipd.it} \\ 
}

\begin{document}
\maketitle

\begin{abstract}
Associative memory models classically describe retrieval as convergence to stored patterns, but the theory of sequential retrieval remains comparatively limited. We develop a two-timescale input-driven Hopfield network in which fast associative retrieval is modulated by a slow feedback variable evolving over a prescribed memory-transition graph. This yields autonomous, graph-constrained transitions while preserving analytical tractability. Using slow-fast and geometric singular perturbation theory, we characterize memory fixation via memory manifolds and identify the geometric regions that govern stability loss and escape from retrieved memories. We further derive reduced slow dynamics and explicit criteria for transition onset, self-sustained retrieval, and collapse, expressed through escape times, gain thresholds, and fixed points. Together, these results provide a tractable framework for analyzing feedback-driven sequential retrieval in Hopfield networks.
\end{abstract}

\section{Introduction}
Associative memory has been a foundational concept in theoretical neuroscience and machine learning for over four decades. In its classical formulation, associative memory retrieval is modeled as convergence toward stable equilibria of an energy landscape~\citep{HJJ:82,CMA-GS:83, HJJ:84}, whereby a corrupted cue relaxes to a stored pattern through recurrent dynamics, providing a canonical example of distributed computation in neural networks. A large body of work has focused on characterizing the memory capacity of such models~\citep{ADJ-GH-SH:87b, MR-PE-RE:87, TMV-FMV:88}, revealing a fundamental limitation of the original Hopfield network, whose capacity scales sub-linearly with the number of neurons. Modern Hopfield models~\citep{KD-HJJ:16,KD-HJJ:20,DM-HJ-LM:17} overcome this limitation by achieving exponential capacity within an energy-based framework, and have been linked to contemporary architectures such as Transformers~\citep{RH-SB-LJ:21,HB-LY-PB:23} and diffusion models~\citep{HB-SH-KD:23,AL:24}. However, they largely remain static retrieval systems: dynamics stop at convergence, so sequential retrieval requires repeated reinitialization. This contrasts with cognitive and algorithmic tasks, including those addressed by Transformers, where reasoning depends on structured transitions across sequences of memories.

Pioneering works by \citep{KD:86,SH-KI:86} already identified the limitation of a theory accounting only for static retrieval. They proposed a continuous-time mechanism for sequential memory transitions via delayed asymmetric interactions, which introduce a destabilizing dynamic component driving transitions between attractors. This idea has since motivated extensive efforts to model memory transitions~\citep{TDW-HJJ:87,KD-SH:88}, establishing mean-field results on the gain sustaining sequentiality. More recently, \citep{KA-ST-SHT:23} introduced a continuous-time, slow-fast energy-based model with delayed asymmetric interactions, while \citep{TC-HZV-JA:24} investigated long-sequence retrieval and capacity in a discrete-time setting. The softmax-based model of \citep{KA-LP-ST-SHT:25} further provides capacity results and escape-time estimates based primarily on energy arguments. Related approaches use slowly evolving parameters to control attractor switching~\citep{HL-SP-XB:23,FM-PC:24}. Our aim is complementary: rather than improving retrieval performance, we focus on developing an analytically tractable model that provides a mechanistic~understanding~of~sequential~memory~transitions.

\begin{figure}[t]
\begin{center}
\includegraphics[scale=0.65]{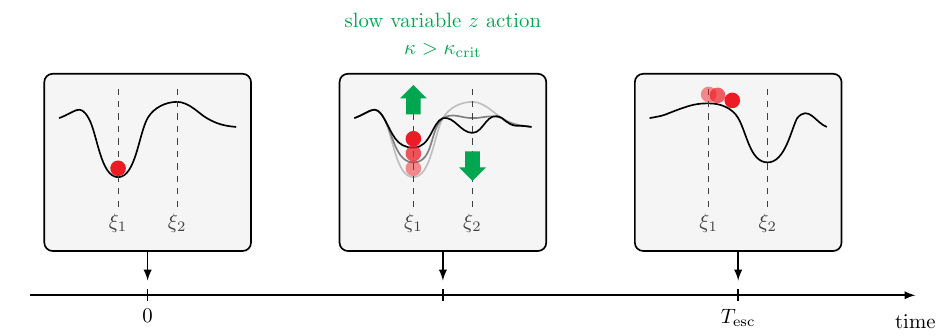}
\end{center}
\caption{\emph{Schematic illustration of the proposed two-timescale mechanism for sequential memory transitions.} 
At the initial time, the fast state variable is at a stable equilibrium corresponding to memory $\upxi_1$. 
The slow variable $z$ progressively destabilizes this equilibrium while simultaneously stabilizing the equilibrium associated with memory $\upxi_2$. At the escape time $T_{\mathrm{esc}}$, the equilibrium corresponding to $\upxi_1$ loses stability and the system transitions toward $\upxi_2$. 
This transition occurs only when the gain parameter $\kappa$ of the slow dynamics exceeds a critical threshold $\kappa_{\mathrm{crit}}$. 
We analytically characterize both $\kappa_{\mathrm{crit}}$ and the resulting escape time $T_{\mathrm{esc}}$.}
\label{fig:intro}
\end{figure}

In this paper, we propose and analyze a two-timescale extension of the input-driven plasticity (IDP) Hopfield model of~\citep{BS-BG-FB-SZ:25idp} that enables autonomous sequential retrieval. In the original IDP Hopfield model, external inputs modulate the saliency of stored memories and thereby reshape the retrieval landscape. Here, we replace this externally prescribed modulation with a slow feedback variable whose dynamics are driven by the currently retrieved memory and structured by a transition graph. The slow feedback can be viewed as a phenomenological representation of activity-dependent biological processes, such as short-term synaptic plasticity, neuronal adaptation, or neuromodulatory gain control, that modulate memory stability~\citep{BMK-FS:16}. The resulting sequential IDP Hopfield network produces graph-constrained transitions between memories while remaining analytically tractable. Our contributions are threefold. First, using slow-fast and geometric singular perturbation theory, we characterize the geometry of memory fixation through memory manifolds and the attracting and saddle-type regions that govern entry and escape. Second, we derive reduced saliency dynamics and explicit escape conditions, including transition times, gain thresholds, and collapse regimes. Third, we show how these conditions support self-sustained sequential retrieval over structured memory graphs. Fig.~\ref{fig:intro} provides a schematic overview of the sequential memory transition mechanism introduced in this work.
Together, these results establish a tractable geometric and dynamical framework for sequential retrieval in input-driven Hopfield networks, providing explicit conditions under which slow internal feedback induces, sustains, or fails to sustain transitions between memories. 

\section{The IDP Hopfield model for sequential retrieval}\label{sec:model}

The IDP Hopfield model introduced in \citep{BS-BG-FB-SZ:25idp} extends the classical Hopfield model by incorporating an explicit modulation of synaptic interactions through external inputs. This formulation enables a controlled and interpretable mechanism for memory retrieval, where the input does not merely select an initial condition but directly shapes the energy landscape governing the dynamics. We next briefly review the IDP Hopfield model, deferring technical details to Appendix~\ref{ssec: IDP_Hop}, and then introduce a sequential extension that enables autonomous, structured transitions between stored memories.

\paragraph{The IDP Hopfield model.}

\begin{figure}[!t]
\begin{center}
\includegraphics[width=.85\linewidth]{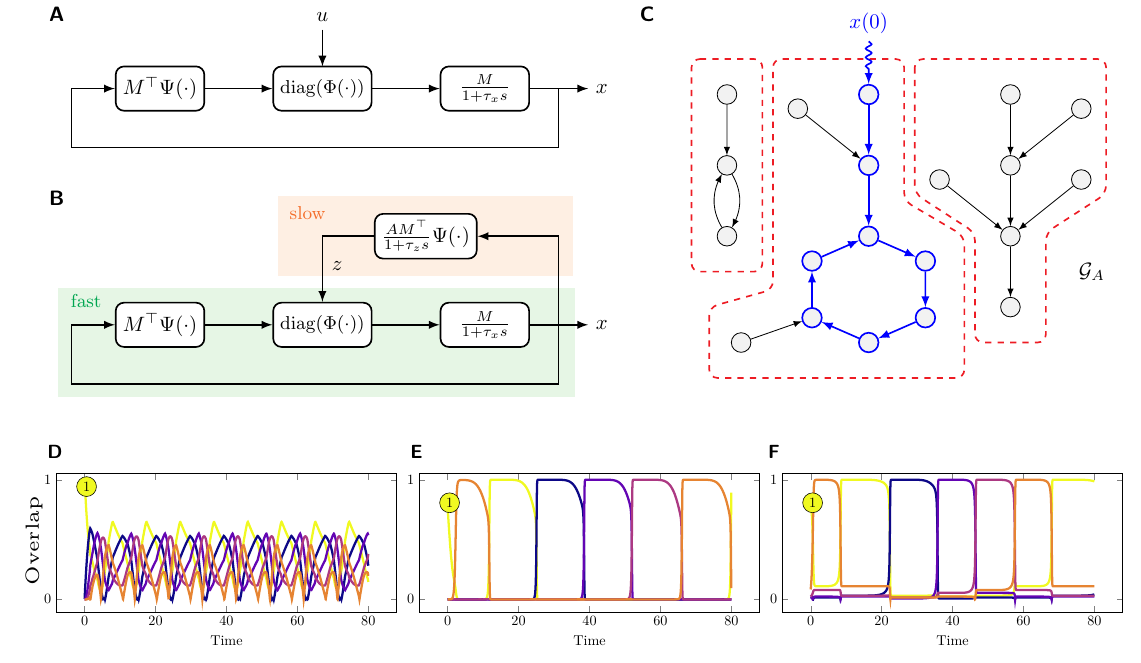}
\end{center}
\caption{
(A) Block diagram of the IDP Hopfield model; setting $\tau_x=1$ recovers (\ref{eq: hopf}). (B) Block diagram of the two-timescale sequential IDP Hopfield model (\ref{eq: ttreas}). 
(C) Example of a deterministic transition graph $\mathcal{G}_A$, where nodes represent stored memories and edges encode admissible transitions. The graph decomposes into multiple connected components (highlighted by dashed red contours), each defining a distinct memory sequence; one, selected by $x(0)$, is highlighted in blue. (D) One-timescale model with orthogonal memories ($\kappa=5$): sequential retrieval proceeds through mixed memory states, with partial overlaps and unpredictable escape times across memories. (E-F) Two-timescale sequential IDP Hopfield model ($\tau_x=1,\tau_z=5,\kappa=5$). (E) Orthogonal memories: transitions occur between individual memories without mixing, with uniform escape times and overlaps reaching their maximal value, indicating exact memory alignment. (F) Random memories: transitions occur without mixing, reaching maximum overlap, but escape times are no longer uniform. Numerical experiments are shown for visualization purposes with $P=15$ memories and a limit cycle over $5$ of them, with $\uppsi=\tanh$, and whose overlaps are highlighted in different colors in the plots.}
\label{fig:block-diagram-IDP}
\end{figure}

The IDP Hopfield dynamics, illustrated schematically in Fig.~\ref{fig:block-diagram-IDP}(A), are
\begin{equation}\label{eq: hopf}
\dot x = -x + W\bigl(\Phi(u)\bigr)\Psi(x), 
\qquad x(0)\in\real^{N},
\end{equation}
where $x\in\real^N$ is the network state and $u\in\real^d$ is a slowly varying external input. 
The map $\Phi:\real^{d}\to\real^{P}_{\geq 0}$ is a \emph{saliency function} that acts as a positive filter on the input $u$, while $\Psi:\real^N\to\real^N$ is the neural \emph{activation function}. The synaptic matrix depends on the filtered input through
\[
W(\Phi(u)) = M \diag(\Phi(u)) M^{\top},
\qquad 
M:=\tfrac{1}{\sqrt{N}}[\upxi^{1},\dots,\upxi^{P}],
\]
where $\{\upxi^{\upmu}\}_{\upmu=1}^{P}\subseteq\{\pm1\}^N$ is a set of mutually orthogonal stored patterns. We stress that the orthogonality constraint is necessary to derive rigorous results on the geometry and dynamics of the discussed model, but numerical experiments support their validity also in the canonical case of random memories.

Throughout the paper, we assume that $\Phi(\cdot)$ acts componentwise and homogeneously, i.e.,
$\Phi_i(u)=\upphi(u_i)$ for some scalar function $\upphi:\real\to\real_{\ge 0}$, which is even and monotone with respect to $|u_i|$. Similarly, $\Psi(\cdot)$ acts componentwise as $\Psi_i(x)=\uppsi(x_i)$, where $\uppsi:\real\to\real$ is odd, continuously-differentiable and monotone non-decreasing (e.g., $\tanh$). Under these assumptions, the map $\Phi(\cdot)$ assigns a non-negative saliency coefficient to each stored pattern, thereby modulating its contribution to the synaptic matrix $W(\Phi(u))$, while $\Psi(\cdot)$ retains the standard symmetry and monotonicity properties of Hopfield-type activation functions. The classical Hopfield model is recovered as a special case when all input-dependent coefficients $\upphi(u_i)$, $i=1,\dots,P$, are all equal to one, in which case the synaptic matrix reduces to a fixed matrix independent of the input.

Under the above-mentioned standard assumptions on $\Psi$, the IDP Hopfield model admits the energy function~\citep{HJJ:84,KD-HJJ:20}
\begin{equation}\label{eq: ener}
\En\bigl(x,\Phi(u)\bigr)
=-\tfrac{1}{2}\Psi(x)^{\top}W\bigl(\Phi(u)\bigr)\Psi(x)
+x^{\top}\Psi(x)
-\sum_{i=1}^{N}\int_{0}^{x_{i}}\uppsi(s)\,ds.
\end{equation}
This energy provides a geometric interpretation of retrieval: the input-dependent saliencies $\Phi(u)$ reshape the energy landscape, thereby controlling the existence and stability of memory equilibria.

A key feature of the IDP formulation is that the input-dependent coefficients $\{\upphi(u_\upmu)\}_{\upmu=1}^{P}$ directly control the existence and stability of memory states. In general, the system admits multiple equilibria, some of which are associated with the stored patterns.\footnote{Other equilibria (so-called \emph{spurious} states) may exist, but they do not play a relevant role in the setting considered here.} In particular, to each memory vector $\upxi^\upnu$ one can associate a \emph{memory equilibrium} of the form $x=\upgamma_\upnu \upxi^\upnu$, which exists if and only if $\upphi(u_\upnu)>1$. In that case, the scaling factor $\upgamma_\upnu$ is uniquely determined by $\upgamma_\upnu=f^{-1}(\upphi(u_\upnu))$, where $f(\upgamma):=\upgamma/\uppsi(\upgamma)$. The stability of these equilibria is also governed by the coefficients $\upphi(u_\upnu)$: if $\upphi(u_\upnu)>1$, the corresponding equilibrium is stable only if $\upphi(u_\upnu)$ is sufficiently large relative to the maximum coefficient $\upphi_{\max}:=\max_{\upmu}\upphi(u_\upmu)$ (see Appendix~\ref{ssec: IDP_Hop} for details). Overall, the coefficients $\upphi(u_\upmu)$ provide a direct and tunable mechanism to determine which memories exist and are stable, thereby enabling selective activation and suppression within the same network.

\paragraph{The Two-Timescale Sequential IDP Hopfield Model.} While the IDP formulation enables input-modulated retrieval and sharp control over memory stability, it remains fundamentally static in nature: for a fixed input, the dynamics converge to a single equilibrium and do not exhibit autonomous transitions between memories. In particular, the model lacks an internal mechanism to encode temporal or semantic relationships among stored patterns. 

To overcome this limitation, we introduce a minimal two-timescale extension of (\ref{eq: hopf}) that couples the fast retrieval dynamics to a slower internal state:
\begin{equation}\label{eq: ttreas}
\begin{cases}
\uptau_x \dot x = -x + M \diag\bigl(\Phi(z) \bigr) M^{\top}\Psi(x), & x(0)\in\real^N,\\
\uptau_z \dot z = -z + \tfrac{\kappa}{\sqrt{N}} A M^{\top}\Psi(x), & z(0)\in\real^P,
\end{cases}
\end{equation}
where $\kappa>0$ is a gain parameter and $A\in\{0,1\}^{P\times P}$ encodes structured associations among memories. Specifically, $A$ is the adjacency matrix of a directed graph $\mathcal{G}_A=(\mathcal{V},\mathcal{E})$, whose nodes $\mathcal{V}=\{1,\dots,P\}$ represent stored memories and whose edges represent admissible transitions, with $(i,j)\in\mathcal{E}$ if and only if $A_{ji}=1$. We focus on deterministic graphs, where each node has out-degree at most one, and assume $\tau_x \ll \tau_z$, so that $x$ evolves rapidly while $z$ provides slow modulation.

Fig.~\ref{fig:block-diagram-IDP}(B) illustrates the block diagram of the sequential IDP Hopfield model (\ref{eq: ttreas}). In contrast to the original IDP formulation (\ref{eq: hopf}), where the input directly modulates the synaptic weights, here the modulation is mediated by the slow, feedback-driven variable $z$, which endows the system with an internal memory of past retrieval states. Fig.~\ref{fig:block-diagram-IDP}(C) shows an example of deterministic graph $\mathcal{G}_A$. Such graphs may consist of multiple connected components, each corresponding to a distinct memory sequence that can be selectively retrieved through appropriate choice of the initial condition $x(0)$.

In the regime $\tau_x \ll \tau_z$, a clear separation of timescales emerges in (\ref{eq: ttreas}): the fast variable $x$ rapidly relaxes toward a memory determined by the current $\Phi(z)$, while the slow variable $z$ evolves in response to the retrieved pattern and progressively reshapes the energy landscape. This interaction gives rise to autonomous sequential retrieval, whereby the system transitions between stored memories according to the graph $\mathcal{G}_A$. As illustrated in Fig.~\ref{fig:block-diagram-IDP}(E-F) on a simple cyclic graph with $P=5$ memories, the proposed two-timescale sequential IDP Hopfield model exhibits robust and predictable sequential retrieval when $\kappa$ is sufficiently large. For orthogonal memories (Fig.~\ref{fig:block-diagram-IDP}(E)), transitions occur directly between individual memories without intermediate mixing: overlaps reach their maximal value, indicating exact memory alignment, and escape times are uniform across memories. For random memories (Fig.~\ref{fig:block-diagram-IDP}(F)), the dynamics similarly transition without mixing and attain maximal overlap, although escape times are no longer uniform. 
To highlight the role of timescale separation, we compare this behavior with the classical sequential Hopfield dynamics obtained in a single-timescale setting, $\dot x = -x + \kappa MAM^\top\Psi(x)$, where transitions are encoded directly in the synaptic matrix~\citep{GM-BN:24}. As shown in Fig.~\ref{fig:block-diagram-IDP}(D), this model exhibits qualitatively different dynamics: trajectories evolve through mixed states with partial overlaps, display irregular escape times, and are highly sensitive to the gain $\kappa$. Depending on its value, the system may collapse to the origin, converge to a fixed point, or sustain a limit cycle, consistent with previous observations.

Finally, we point out that (\ref{eq: ttreas}) admits a natural modern Hopfield reformulation, detailed in Appendix~\ref{ssec:modern_IDP}.

\section{Memory manifolds: the highway of sequential retrieval}\label{sec:GSPT}
\emph{Sequential memory retrieval} is the continuous evolution of neural activity in state space, from one stable memory state to another one. The direction of motion is shaped by the semantic relationships between memories, which at any given time constrain neural activity to a lower-dimensional memory subspace. As semantic relationships change, the binding is broken, allowing neural trajectories to leave one memory subspace and evolve toward another.

\begin{figure}[!h]
    \centering
    \includegraphics[width=.8\linewidth]{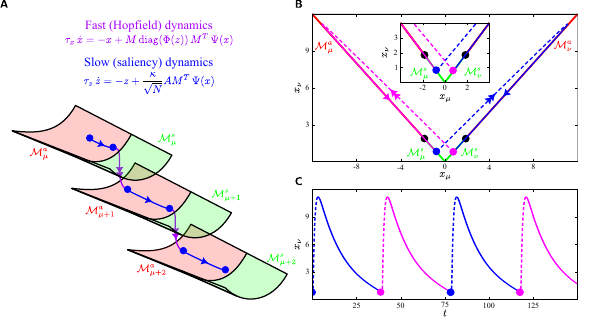}
    \caption{(A) Sketch of the transitions between different memory manifolds in the slow-fast Hopfield model. The blue curves corresponds to the slow movement of the system close to a stable memory manifold, whereas the purple curves correspond to the fast switching between different stable memory manifolds. 
    Each memory manifold $\mathcal{N}_\upmu$ is decomposed into its attracting (red) and saddle-type (green) memory manifolds $\mathcal{M}_{\upmu}^a$, $\mathcal{M}_{\upmu}^s$. 
    (B) Simulated trajectory projected onto the fast subspace in a slow-fast Hopfield model storing two memories. 
    Continuous curves indicate the flow along memory branches $\mathcal{N}_\upmu$ (magenta) and $\mathcal{N}_\upnu$ (blue), whereas dashed curves indicate rapid transitions from one memory branch to another. 
    Switching occurs after the trajectory has moved from the attracting part of the memory branches $\mathcal{M}_{\upnu}^{a,\varepsilon}$ and $\mathcal{M}_{\upmu}^{a,\varepsilon}$, respectively, to the saddle-type section $\mathcal{M}_{\upnu}^{s,\varepsilon}$ and $\mathcal{M}_{\upmu}^{s,\varepsilon}$.
    The black dots indicate the relevant point of separation between $\mathcal{M}_{\upnu}^{a}$ and $\mathcal{M}_{\upnu}^{s}$, and $\mathcal{M}_{\upmu}^{a}$ and $\mathcal{M}_{\upmu}^{s}$, respectively (red/green).
    (C) Temporal series of the variable $x_\upnu$. Colors and line types as in panel B.}
    \label{fig: manifolds}
\end{figure}

From a geometric perspective, as sketched in Fig.~\ref{fig: manifolds}(A), sequential retrieval therefore requires two complementary structures: (i) \emph{memory subspaces} (red and shaded surfaces) that transiently trap neural activity (blue curves), and (ii) \emph{entry and exit pathways} that enable transitions (purple branches) between them. In this section, we formalize this intuition using slow-fast and \emph{geometric singular perturbation theory} (GSPT)~\citep{FN:79,HG:09}. Leveraging the analytical tractability of the IDP Hopfield model, we provide a geometric characterization of how sequential retrieval arises in high-dimensional associative memory systems.

Geometric singular perturbation theory~\citep{FN:79} provides a
theoretical framework to analyze dynamical systems with well-separated time scales.
The central idea is that, when fast and slow variables interact weakly, the full dynamics can be approximated as a sequence of slow movements, solutions of a simpler system, interspersed by fast jumps. We introduce the time-scale parameter $\varepsilon=\uptau_x/\uptau_z$ and consider the slow-fast system
\begin{equation}\label{eq: gspt-HOP}
\begin{cases}
\varepsilon \dot x = -x+M\,\diag(\Phi(z))\,M^{\top}\Psi(x), \quad\:\:\:\: x(0)\in\mathbb{R}^N,\\
\dot z = -z + \tfrac{\kappa}{\sqrt{N}}\, A M^{\top}\Psi(x), \qquad\qquad\quad\ \ z(0)\in\mathbb{R}^P.
\end{cases}
\end{equation}
We refer to the $x$-ODE with the term \emph{fast subsystem} and denote with $F(x,z)$ its right-hand side.
Taking the limit $\varepsilon\to0^+$ constrains the fast dynamics onto $\vectorzeros[N] = F(x,z)$, yielding the \emph{reduced system}.
The algebraic constraint effectively tracks the minimum over the energy as the slow neural feedback evolves, and admits multiple solutions, reflecting that Hopfield networks are designed to store multiple stable patterns. In slow-fast theory, the set of the fast-subsystem equilibria is called \emph{critical set}, and defined as
\begin{equation}
    \mathcal{S}_0 = \left\{(x,z) \in \mathbb{R}^N \times \mathbb{R}^P: \vectorzeros[N] = F(x,z) \right\}.
\end{equation}

\paragraph{Memory branches, memory manifolds, and stability.}
Given the fast dynamics with $P$ stored memories, a \emph{memory branch} $\mathcal{N}_\upnu \subseteq \mathcal{S}_0$ describes a $P$-dimensional subspace in which neural activity can be temporarily confined. In the IDP Hopfield model, these memory branches~take~the~explicit~form
\begin{equation}\label{set: IDPeq}
\mathcal{N}_{\upnu}
=\left\{(x,z)\in\mathbb{R}^N\times\mathbb{R}^P :\ \exists\upgamma_\upnu\neq 0,\:  
x=\upgamma_\upnu \upxi^\upnu,\ 
\upphi(z_\upnu)=f(\upgamma_\upnu)
\right\},
\end{equation}
for $\upnu=1,\dots,P$ and $f(\upgamma)=\upgamma/\uppsi(\upgamma)$. The \emph{normally hyperbolic} subsets $\mathcal{M}_\upnu^{a}$ and $\mathcal{M}_\upnu^s$ of $\mathcal{N}_\upnu$ are known as the attracting and saddle-type \emph{memory manifolds}. They can be identified by studying the stability of the fast subsystem equilibria in $\mathcal{N}_{\upnu}$, which are related to the saliency weights.
Another relevant \emph{branch} of $\mathcal{S}_0$ is the \emph{origin manifold},
\begin{equation}
    \mathcal{M}_0 = \left\{(x,z) \in \mathbb{R}^N \times \mathbb{R}^P : x = 0\right\}.
\end{equation}
GSPT ensures that, for $0<\varepsilon\ll1$, each $\mathcal{N}_{\upnu}$ ($\mathcal{M}_{\upnu}^{a}$, $\mathcal{M}_\upnu^s$, and $\mathcal{M}_0$) perturbs to a nearby invariant manifold $\mathcal{N}_{\upnu}^\varepsilon$ ($\mathcal{M}_{\upnu}^{a,\varepsilon}$, $\mathcal{M}_\upnu^{s,\varepsilon}$, and $\mathcal{M}_0^\varepsilon$) of the full system~(\ref{eq: gspt-HOP}). In particular, the manifolds $\mathcal{M}_\upnu^{a,\varepsilon}$ and $\mathcal{M}_\upnu^{s,\varepsilon}$ provide a precise geometric description of \emph{memory fixation} and \emph{transitions}.

\paragraph{Fixation, loss of stability, and transitions.}
Sequential retrieval requires understanding how trajectories \emph{enter} and \emph{leave} $\mathcal{N}_\upnu^\varepsilon$, which consists in studying the stability of its constitutive memory manifolds. As presented in Fig.~\ref{fig: manifolds}(B-C), during \emph{retrieval} of the memory $\upxi^\upnu$, the neural activity converges to $\mathcal{M}_\upnu^{a,\varepsilon}$ rapidly, fixation occurs, and escape is impossible. As the saliency $\upphi(z_\upnu)$ decreases, an escape direction appears when $\upphi(z_\upnu)$ crosses the threshold $\upphi_{\max}$ from above. Geometrically, this corresponds to crossing (an $\varepsilon$-perturbation of) a non-hyperbolic point on $\mathcal{N}_\upnu$ (black circle in Fig.~\ref{fig: manifolds}(B)), corresponding to the trajectory moving from the attracting manifold $\mathcal{M}_\upnu^{a,\varepsilon}$ to the saddle manifold $\mathcal{M}_{\upnu}^{s,\varepsilon}$. Beyond this point, a typical trajectory is ejected via an unstable fiber of $\mathcal{M}_\upnu^{s,\varepsilon}$ towards a new attracting manifold $\mathcal{M}_\upmu^{a,\varepsilon}$ associated to the memory $\upxi^\upmu$. However, the jump towards a different memory can occur only if the associated $\upphi(z_\upmu) >  1$. Otherwise, if for all $\upmu$ it holds $\upphi(z_\upmu) < 1$ then the only attracting manifold is $\mathcal{M}_0^\varepsilon$, which drives the full system towards the origin.

\section{Slow saliency dynamics: the driving engine}\label{sec:slow-reasoning}

In Sec.~\ref{sec:GSPT}, we characterized the geometric structure underlying memory retrieval, including the manifolds that support retrieval and the instabilities and fibers that enable transitions between them. We now turn to how the reduced slow dynamics in~(\ref{eq: gspt-HOP}) govern sequential transitions as a function of the gain parameter $\kappa$, the initial condition $z(0)$, and the structure of the transition matrix $A$. We present the main results below, deferring technical derivations and proofs to Appendix~\ref{app:slow-reas}.

\paragraph{Reduced dynamics and retrieval regimes.}
Consider the sequential IDP Hopfield model~(\ref{eq: ttreas}) in the limit
$\tau_x/\tau_z\to 0^+$. The fast variable $x$ rapidly relaxes to a stable equilibrium $x^*(z)$ determined by the current saliency $z$. Substitution into the slow~equation~of~(\ref{eq: ttreas})~gives
\begin{equation}\label{eq: GSPT-slow}
\dot z = -z + \frac{\kappa}{\sqrt{N}}\, A M^{\top}\Psi\bigl(x^*(z)\bigr).
\end{equation}
This reduced equation describes the slow flow along the branches $\{\mathcal{N}_\upmu\}_{\upmu=1}^P$ of the fast Hopfield subsystem: for each fixed $z$, the fast dynamics select an equilibrium $x_\upnu^*(z)$ and sticks to the associated attracting manifold $\mathcal{M}_\upnu^a$ until it loses stability. Only then the trajectory may jump to another memory branch, as pictorially depicted in the introductory Fig.~\ref{fig:intro}. 

By the geometric picture of Sec.~\ref{sec:GSPT}, the saliency state $z$ selects which memory branches contain attracting manifolds and can therefore support fixation. This gives three regimes:
\begin{itemize}
\item[(i)] \textbf{No active memories.}
If $\upphi(z_\upmu)<1$ for all $\upmu$, no memory equilibrium exists and trajectories converge to the origin by evolving in $\mathcal{M}_0$.

\item[(ii)] \textbf{Unique active memory.}
If there is a unique $\upnu$ such that $\upphi(z_\upnu)>1$, there is only active memory equilibrium, and trajectories rapidly converge to $\mathcal{M}_\upnu^a$. This equilibrium is stable and given by
$$
x^*(z)=f^{-1}\bigl(\upphi(z_\upnu)\bigr)\,\upxi^\upnu, \ \ \text{ where } f(\upgamma) = \upgamma/\uppsi(\upgamma).
$$

\item[(iii)] \textbf{Competing memories.}
If several components $\upphi(z_\upnu)$ exceed the threshold, multiple memory equilibria may coexist, allowing transitions between the corresponding memory states.
\end{itemize}

\paragraph{Two-memory reduction.}
We next restrict to the case in which at most two memories compete. Specifically, we assume that, along the relevant portion of the trajectory, only two components of $z(t)$ are significantly different from zero. Without loss of generality, let the initial time~be~$t=0$~and~assume $z(0)=Z\ve^{(\upnu)},$ where $\ve^{(\ell)}$ denotes the $\ell$-th vector of the canonical basis of $\mathbb{R}^P$ and $Z>0$ is the initial value of $z_\upnu$.\footnote{We take $Z>0$ for simplicity. The analysis can be easily extended to $Z\in\mathbb{R}$. We note also that if $Z$ were negative, the transition update below yields $Z^+>0$, and all subsequent iterations have non-negative amplitudes.} 

If $\upphi(Z)<1$, no memory is active: the fast state converges to $x(t)=0$, and the slow dynamics reduce to $\dot z=-z$, driving all saliencies exponentially to zero in $\mathcal{M}_0$. We therefore focus on $\upphi(Z)>1$, where $\upxi^\upnu$ is initially active and the fast state is trapped on its associated memory equilibrium,
$$
x^\star_\upnu(t)=f^{-1}\bigl(\upphi(z_\upnu(t))\bigr)\,\upxi^\upnu.
$$
In this case, the system initially evolves on $\mathcal{M}_\upnu^a$, and the reduced saliency dynamics (\ref{eq: GSPT-slow}) satisfy
\begin{align}
    \dot z
    &= -z+\kappa\,
    \uppsi\left(f^{-1}\bigl(\upphi(z_\upnu)\bigr)\right)
    \ve^{(\upmu)},
\end{align}
where $\upmu$ is the unique successor of $\upnu$ in the transition graph $\mathcal{G}_A$. Hence $\dot z_\upsigma=-z_\upsigma$ for all $\upsigma\neq\upmu$. Since only $z_\upnu(0)$ is nonzero, all components except $z_\upnu$ and $z_\upmu$ remain identically zero, with $z_\upnu(t)=Ze^{-t}$. The successor saliency component satisfies
$$
\dot z_\upmu=-z_\upmu+\kappa\,g\bigl(Ze^{-t}\bigr), \qquad g:=\uppsi\circ f^{-1}\circ\upphi .
$$
Note that, on the interval where the source memory exists, $g(Ze^{-t})>0$, so $z_\upmu$ is driven by a positive input. Thus, the transition from $\upxi^\upnu$ to $\upxi^\upmu$ reduces to a two-dimensional saliency dynamics: a decaying source component and a positively driven successor component.

\paragraph{Transition criterion and induced map.}
We now determine whether the two-dimensional saliency dynamics produces a transition to the successor memory or decays to inactivity. Let $T_{\max}$ be the time at which the source memory ceases to exist, namely $\upphi\bigl(z_\upnu(T_{\max})\bigr)=1$. Since $z_\upnu(t)=Ze^{-t}$, the source memory exists for $t<T_{\max}$ and disappears for $t>T_{\max}$. The outcome is determined by whether the successor saliency $z_\upmu(t)$ becomes large enough before this loss of existence occurs. Two possible cases arise (see also Fig.~\ref{fig:zampi} of the Appendix for a graphical illustration).

\emph{Case 1: no transition.}
Assume that
$$
z_\upmu(t)<z_\upnu(t)
\qquad \forall\,t\in[0,T_{\max}).
$$
Then the successor component never catches up with the source component before the source memory disappears. Hence, after $T_{\max}$, both saliencies fall below the existence threshold and the system enters the inactive regime $\mathcal{M}_0$, where the origin is the only stable equilibrium.

\emph{Case 2: transition.}
Assume instead that there exists a first time $\bar T\in[0,T_{\max})$ such that
$$
z_\upmu(\bar T)=z_\upnu(\bar T).
$$
In this case, the following result characterizes the loss of stability of the source branch from $\mathcal{M}_\upnu^a$ to $\mathcal{M}_\upnu^s$, and the induced update of the successor saliency. The proof is given in Appendix~\ref{app:slow-reas}.

\begin{theorem}[Escape condition and saliency update]\label{thm:escape-time}
Under the above condition, there exists an escape time $T_{\mathrm{esc}}\in[\bar T,T_{\max})$ at which the equilibrium of the fast Hopfield subsystem associated with the source memory $\upxi^\upnu$ loses stability. Moreover, $T_{\mathrm{esc}}$ is the smallest solution in $[0,T_{\max})$ of
\begin{equation}\label{eq:escape-time}
\kappa\int_0^{T_{\mathrm{esc}}} e^{-(T_{\mathrm{esc}}-\tau)}
g\bigl(Ze^{-\tau}\bigr)\,d\tau
=
h\bigl(Ze^{-T_{\mathrm{esc}}}\bigr),
\end{equation}
where $h(z):=\upphi^{-1}\!\left(\tfrac{1}{\uppsi'\!\left(f^{-1}(\upphi(z))\right)}\right)$.\footnote{Here $\upphi^{-1}$ is taken on $\real_{\geq 0}$, where $\upphi$ is invertible by assumption.} At escape, the successor saliency is
\begin{equation}\label{eq:Zplus-escape}
Z^+:=z_\upmu(T_{\mathrm{esc}})
=
h\bigl(Ze^{-T_{\mathrm{esc}}}\bigr).
\end{equation}
\end{theorem}

At time $T_{\mathrm{esc}}$, the trajectory leaves the branch associated with the source memory $\upxi^\upnu$ and jumps to the manifold $\mathcal{M}_\upmu^a$ associated with the successor memory $\upxi^\upmu$. Theorem~\ref{thm:escape-time} therefore defines a one-dimensional saliency update $Z\mapsto Z^+$.

For later use, it is convenient to rewrite this update in terms of $X:=e^{T_{\mathrm{esc}}}$. Since
$$\int_0^{T_{\mathrm{esc}}} e^{-(T_{\mathrm{esc}}-\tau)}g\bigl(Ze^{-\tau}\bigr)\,d\tau=\frac{1}{X}\int_1^X g\left(\frac{Z}{s}\right)\,ds,$$
the transition map is equivalently described by\footnote{The range constraint $X\in[1,Z/\upphi^{-1}(1))$ follows from
$T_{\mathrm{esc}}\in[0,T_{\max})$. In fact, since $z_\upnu(t)=Ze^{-t}$ and
$\upphi(Ze^{-T_{\max}})=1$, we have
$e^{T_{\max}}=Z/\upphi^{-1}(1)$.}
\begin{equation}\label{eq:saliency-map}
\begin{cases}
R(Z,X)=0,\\[1mm]
Z^+=h(Z/X),
\end{cases}
\qquad
X\in[1,Z/\upphi^{-1}(1)),
\end{equation}
where
\begin{equation}\label{eq:F-map}
R(Z,X):= \frac{\kappa}{X}\int_1^X g\left(\frac{Z}{s}\right)\,ds-h\left(\frac{Z}{X}\right).
\end{equation}
Thus, each saliency update is computed by first selecting the smallest 
$X\in[1,Z/\upphi^{-1}(1))$ satisfying $R(Z,X)=0$, when such a solution exists, and then setting
$Z^+=h(Z/X)$. If no such $X$ exists, the transition fails and the trajectory collapses to the inactive state.

\section{Analytical illustration and numerical validation}

We illustrate the results of Sec.~\ref{sec:slow-reasoning} with an analytically tractable example and further assess their validity and robustness through numerical experiments. See also Appendix~\ref{app:exp}.

We consider the saliency update (\ref{eq:saliency-map})-(\ref{eq:F-map}) with the following nonlinearities
\begin{equation}\label{eq:smoothf}
\uppsi(x)=\frac{x}{1+|x|}, \qquad \upphi(z)=|z|.
\end{equation}
In particular, we will refer to such $\uppsi$ as SmoothF. On the positive branch $z\ge 1$, we have
$f(\gamma)=\gamma/\uppsi(\gamma)=\gamma+1$, hence $f^{-1}(z)=z-1$. Therefore
$g(z)=\uppsi(f^{-1}(\upphi(z)))=1-1/z$ and $h(z)=z^2$ for $z\ge 1$. Substituting into
(\ref{eq:F-map}), for $Z>1$ and $X\in[1,Z)$,
$$
\int_1^X g\left(\frac{Z}{s}\right)\,ds =(X-1)\left(1-\frac{X+1}{2Z}\right).
$$
Hence the equation $R(Z,X)=0$ is equivalent to
\[
G(Z,X)=\frac{1}{\kappa}, \qquad G(Z,X):=-\frac{X(X-1)(X+1-2Z)}{2Z^3}.
\]
Thus, given $Z$, the update is computed by first selecting the smallest admissible
solution $X\in[1,Z)$ of this equation, and then setting $Z^+=Z^2/X^2$. If no
such solution exists, the transition fails.

Fixed points of this scalar map characterize self-sustained retrieval across repeated transitions: $Z^+=Z$ means that each jump regenerates the same saliency amplitude. Since $Z^+=Z^2/X^2$, fixed points satisfy $Z=X^2$, and the implicit equation for $X$ becomes
$$
\frac{1}{\kappa}= G(X^2,X)=\frac{(X-1)^2(2X+1)}{2X^5}.
$$
Thus fixed points exist when the level $1/\kappa$ intersects the curve $G$. Since $G$ has a unique maximum at $\bar X=(5+\sqrt{105})/8$, with $G(\bar X^2,\bar X)\simeq 0.0785$, fixed points exist only for $\kappa\ge\kappa_{\mathrm{crit}}:=1/G(\bar X^2,\bar X)\simeq 12.74$. In particular, for
$\kappa>\kappa_{\mathrm{crit}}$, two fixed points appear $\bar Z_1$ and $\bar Z_2$, $\bar Z_1<\bar Z_2$, with $\bar Z_1$ unstable and $\bar Z_2$ stable. Since $X=e^{T_{\mathrm{esc}}}$ and $\bar Z_2=X^2$ at the stable fixed point, the asymptotic escape time is $T_{\mathrm{esc}}^\infty=\frac12\log\bar Z_2$. Further details are given in Appendix~\ref{app:slow-reas}.

Fig.~\ref{fig: z_strength} illustrates the preceding analysis and extends it numerically to the case $\uppsi=\tanh$. Fig.~\ref{fig: z_strength}(B) compares the SmoothF and Tanh activation functions, while Fig.~\ref{fig: z_strength}(A) shows the induced saliency update map for SmoothF, revealing the emergence of two fixed points for gains exceeding the critical value $\kappa>\kappa_{\mathrm{crit}}\simeq12.74$. Fig.~\ref{fig: z_strength}(C)-(F) present representative trajectories for subcritical and supercritical values of $\kappa$ using the SmoothF activation. For orthogonal memories (Fig.~\ref{fig: z_strength}(C)-(D)), subcritical gains lead to collapse into an inactive state, whereas supercritical gains give rise to self-sustained sequential retrieval. In contrast, for random memories (Fig.~\ref{fig: z_strength}(E)-(F)), both gain regimes exhibit sustained transitions, likely due to cross-talk noise and the consequent lack of entrainment to a fixed asymptotic value. In the subcritical regime, the resulting $z$-oscillations have small amplitude and may therefore lead to collapse only on longer timescales. Finally, Fig.~\ref{fig: z_strength}(G)-(H) show that the qualitative features of the slow $z$-dynamics persist persist with Tanh activation and random memories.

\begin{figure}[!t]
    \centering
    \includegraphics[width=0.9\linewidth]{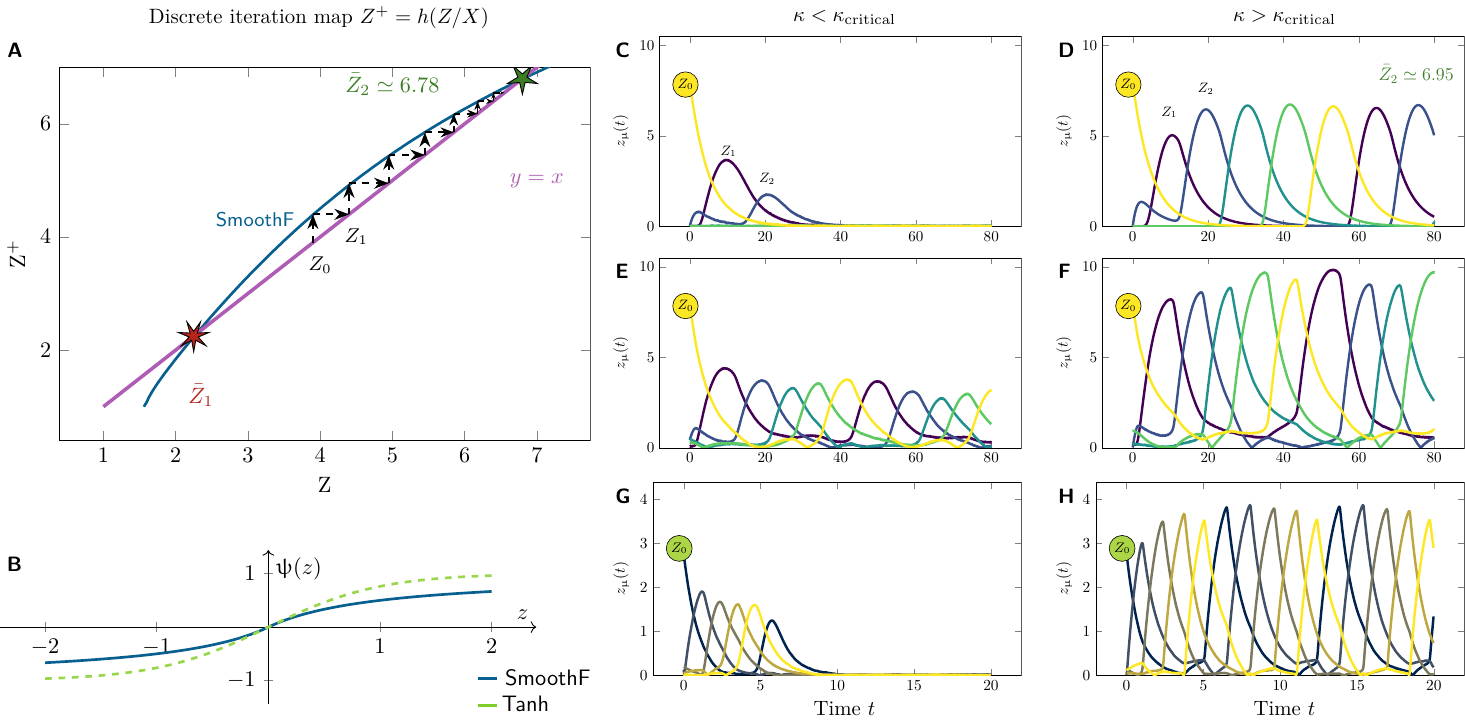}
    \caption{
    (A) Fixed points of the discrete map $Z^+=h(Z/X)$ with $\kappa=15$ for the SmoothF activation, with predicted $\bar Z_2\simeq 6.78$ (see Fig.~\ref{fig:tanh-app} in Appendix~\ref{app:tanh} for $\uppsi=\tanh$). Real solutions $\bar{Z}_1,\bar{Z}_2>1$ emerge only for $\kappa \geq \kappa_{\mathrm{crit}}\simeq12.74$; self-sustained transitions require $Z_0 > \bar{Z}_1$. (B) SmoothF (blue) and Tanh (green) activation functions. (C, D) Orthogonal memories with SmoothF activation. (C) Subcritical gain ($\kappa=9$): oscillations decay and retrieval collapses to the origin. (D) Supercritical gain ($\kappa=15$): persistent oscillations converge to the stable fixed point $\bar{Z}_2\simeq 6.95$, close to theoretical prediction, enabling reliable sequential retrieval. (E, F) Random memories with SmoothF activation. (E) Subcritical gain: cross-talk noise sustains small-amplitude transitions. (F) Supercritical gain: transitions persist but exhibit amplitude fluctuations without entrainment to $\bar{Z}_2$. (G,H) Random memories with Tanh activation. (G) Subcritical gain ($\kappa=3$) leads to collapse, with slower decay due to cross-talk. (H) Supercritical gain ($\kappa=5$) yields self-sustained transitions. Panels (C-H) use $N=256$, $P=15$, and a directed cycle of length $L=5$.}
    \label{fig: z_strength}
\end{figure}

\begin{figure}
    \centering
    \includegraphics[width=\linewidth]{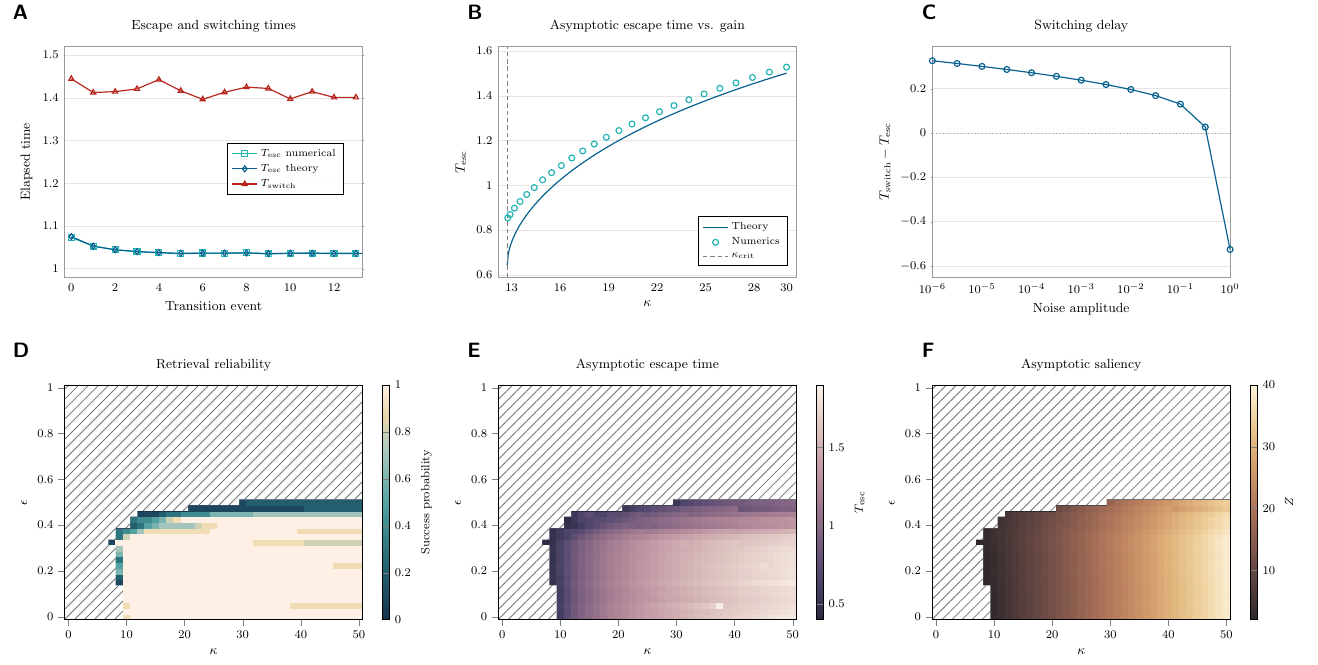}
    \caption{
    (A) Theoretical and numerical escape times $T_{\mathrm{esc}}$ and observed switching times $T_{\mathrm{switch}}$ across consecutive transitions ($\kappa=15$, $Z_0=8$).
    (B) Asymptotic escape time versus $\kappa$: theory and numerical estimates.
    (C) Mean switching delay $T_{\mathrm{switch}}-T_{\mathrm{esc}}$ versus Gaussian noise amplitude $\sigma$.
    (D) Probability of correct transitions in the $(\kappa,\epsilon)$ plane.
    (E) Asymptotic escape time versus $\kappa$ and memory bias $\epsilon$.
    (F) Asymptotic saliency amplitude versus $\kappa$ and $\epsilon$.
    The numerical configuration is the same as in Fig.~\ref{fig: z_strength}. Panels (A-C) use orthogonal memories, while (D-F) use biased random memories, with 10 independent realizations per parameter pair. In (B), asymptotic escape times are estimated over the last 8 time units of simulations of duration $T=40$ (approximately 5-7 transitions). Panel (C) uses Gaussian noise with standard deviation $\sigma/\sqrt{N}$ and 64 realizations per noise amplitude. Hatched regions in (D-F) indicate no successful transitions.
    }
    \label{fig:corr_main}
\end{figure}

We next validate the theoretical escape-time predictions through simulations of the full model. For orthogonal memories, numerical escape times closely match the predictions (Fig.~\ref{fig:corr_main}(A)), while the switching times $T_{\mathrm{switch}}$, defined as the times at which the network transitions to the successor memory, occur slightly later due to the finite time required to leave the unstable memory state. Agreement between theoretical and numerical asymptotic escape times $T_{\mathrm{esc}}^\infty=\frac12\log\bar Z_2$ improves with $\kappa$ (Fig.~\ref{fig:corr_main}(B)). Adding Gaussian noise to the fast dynamics reduces the mean switching delay $T_{\mathrm{switch}}-T_{\mathrm{esc}}$, eventually making it negative due to premature switching (Fig.~\ref{fig:corr_main}(C)).

We then investigate robustness to non-orthogonal memories by independently sampling biased binary patterns, as in~\citep{TC-HZV-JA:24}, according to $\mathbb{P}(\upxi_i^\upmu=\pm1)=(1\pm\epsilon)/2$, where $\epsilon\in[0,1]$ induces an expected normalized pairwise overlap of $\epsilon^2$. Fig.~\ref{fig:corr_main}(D) shows that the critical gain for successful retrieval increases only slightly with memory bias $\epsilon$, while retrieval reliability decreases sharply for $\epsilon\gtrsim0.4$. No successful transitions are observed for $\epsilon>0.5$ over the investigated gains. Conditional on successful transitions, both the asymptotic escape time (Fig.~\ref{fig:corr_main}(E)) and saliency amplitude (Fig.~\ref{fig:corr_main}(F)) increase with $\kappa$. As memory bias $\epsilon$ increases, escape times decrease, while saliency remains approximately constant. Overall, these results suggest that the theoretical estimates derived for orthogonal memories remain informative for moderately correlated memories ($\epsilon\lesssim0.35$), while stronger correlations lead to significant deviations and eventually compromise sequential retrieval. Additional numerical results for the Tanh activation are provided in Appendix~\ref{app:tanh}.

\section{Limitations and future research}
While this work presents an analytically tractable framework for sequential retrieval in Hopfield networks, several important limitations remain. First, our analysis is restricted to canonical Tanh-type activation functions and does not address the practically relevant case $\Psi=\mathrm{softmax}$. This regime is of particular interest, as it directly links Hopfield dynamics to self-attention mechanisms in Transformer architectures~\citep{RH-SB-LJ:21} and is known to enable exponential memory capacity~\citep{DM-HJ-LM:17}. Extending the present theory to non-diagonal activation functions represents a promising direction for future work, with the potential to provide dynamical interpretations of inference in modern Transformer-based models. Second, although we provide a qualitative geometric characterization of the invariant sets and unstable structures supporting sequential retrieval, we do not yet precisely quantify the ranges of slow-fast timescale separation for which transitions emerge intrinsically, without reliance on numerical error or small-amplitude noise to trigger adherence to unstable fibers. A more refined geometric analysis is needed to delineate the parameter regimes that permit robust, noise-free sequential transitions. Third, while our results establish conditions for sequential retrieval, they do not characterize how retrieval reliability scales with the number of stored memories and the length of the retrieved sequence. Quantifying these capacity limits remains an important open problem, preliminarily explored numerically in Appendix~\ref{app:capacity}. Fourth, our theoretical analysis is restricted to deterministic transition graphs with at most one outgoing edge per memory. Extending the theory to branching graphs is another promising direction, as preliminary simulations in Appendix~\ref{app:branching} suggest that memory overlap can influence branch selection. Finally, the role of external inputs in controlling retrieval sequences remains largely unexplored. Developing a theoretical framework for steering memory transitions through external signals is an interesting direction for future research, illustrated by the associative-memory finite-state machine in Appendix~\ref{app:input-ex}.

{\small
\bibliographystyle{unsrtnat}
\bibliography{SB-main}
}


\newpage

\appendix

\appendix
\textbf{\Large Appendix}
\section{The IDP Hopfield model}\label{ssec: IDP_Hop}
The IDP Hopfield model presented in~\cite{BS-BG-FB-SZ:25idp} is characterized by the O.D.E.
\begin{equation}\label{eq: IDP}
    \dot x = -x + W(\upalpha)\Psi(x)\qquad \upalpha\in\real^P_{>0}
\end{equation}
where $W(\upalpha)$ is the synaptic matrix of neural interaction built from the memories of the model. The memories $\upxi^{1},\dots,\upxi^{P}$ and associated synaptic matrix $\W(\upalpha)$ obey the following assumption.
\begin{assumption}[\textbf{Memory and synaptic matrix}]\label{ass: W}
Let $P\in\mathbb{N}$ and define the set of orthogonal memories
\begin{align}\label{eq: mem}
    \Sigma &:=\{\upxi^{1},\dots,\upxi^{P}\}\subseteq\{-1,1\}^{N}\\
    {\upxi^{\upmu}}^{\top}\upxi^{\upnu}&=N\updelta_{\upmu,\upnu}\qquad \upmu,\upnu\in\{1,\dots,P\}
\end{align}
where $\updelta_{\upmu,\upnu}$ is the Kronecker delta for indices $\upmu$ and $\upnu$. For $\upalpha\in\real^{P}_{>0}$ positive we define the input-modulated synaptic matrix $\W(\upalpha)\in\real^{N\times N}$ as
\begin{equation}\label{eq: W(a)}
    \W(\upalpha)=\frac{1}{N}\sum_{\upmu=1}^{P}\upalpha_{\upmu}\upxi^{\upmu}{\upxi^{\upmu}}^{\top}.
\end{equation}
\end{assumption}
The orthogonal memories constraint is necessary to derive rigorous results on existence and stability of fixed points, but can be relaxed to orthogonality in expectation in implementations:
\begin{equation}
    \mathbb{P}[\upxi^{\upnu}_{i}=1]=\mathbb{P}[\upxi^{\upnu}_{i}=-1]=\tfrac{1}{2}\qquad \forall\upnu=1,\dots,P,\quad \forall i=1,\dots,N.
\end{equation}
\begin{remark}
    Notice that by constructing the memory matrix
    \begin{equation}
        M=\tfrac{1}{\sqrt{N}}[\upxi^{1},\dots,\upxi^{P}]
    \end{equation}
    we can rewrite the input-modulated synaptic matrix as
    \begin{equation}
        W(\upalpha)=M\diag(\upalpha)M^{\top}
    \end{equation}
    The canonical synaptic matrix $W$ of the Hopfield model is a specific instance of~(\ref{eq: W(a)}) for the choice $\upalpha=\vectorones$ (or a scalar multiple of it).
\end{remark}
We now introduce an assumption on the saliency function $\Phi$ going from the space of the slow  neural feedback $z\in\real^P$ to the saliency space $\upalpha\in\real^P$ such that $\upalpha=\Phi(z)$.
\begin{assumption}\label{ass: input_filt}
    The saliency function $\Phi:\real^P\to\real^P$ satisfies the following properties:
    \begin{itemize}
        \item[(i)] \textbf{(diagonality and homogeneity)}  for all $\upnu=1,\dots,P$ we have $\Phi_{\upnu}(z)=\upphi(z_{\upnu})$;
        \item[(ii)] \textbf{(symmetry and positivity)}  $\upphi(s)=\upphi(-s)\geq 0$, with $\upalpha_{\upnu}(0)=0$;
        \item[(iii)] \textbf{(symmetric monotonicity)}  $\upphi(s)$ is monotonically increasing on $(0,+\infty)$. 
    \end{itemize}
\end{assumption}

Likewise, we introduce some smoothness assumptions on the activation function $\Psi$ of the IDP Hopfield network, which will be useful for the characterization of the existence and local stability conditions for the equilibria of~(\ref{eq: IDP}).
\begin{assumption}\label{ass: act_f}
    The activation function $\Psi:\real^N\to\real^N$ satisfies the following properties:
    \begin{itemize}
        \item[(i)] \textbf{(diagonality, homogeneity and smoothness)}  for all $i=1,\dots,N$, we have $\Psi_i(x)=\uppsi(x_i)$ with $\uppsi\in C^1(\real)$;
        \item[(ii)] \textbf{(odd and saturating)}  $\uppsi(s)=-\uppsi(-s)$ for all $s\in\real$ and $\lim_{s\to\pm \infty}\uppsi(s)=\pm 1$.
        \item[(iii)] \textbf{(bounded monotonicity)}  $\uppsi'(s)\in(0,1]$ for all $s\in\real$ and $\uppsi'(0)=1$;
        \item[(iv)] \textbf{(monotonicity of the derivative)}  $\uppsi'$ is monotonically decreasing on $(0,+\infty)$.
    \end{itemize}
\end{assumption}
We will now state some of the main analytical results presented in~\cite[Supplementary Information]{BS-BG-FB-SZ:25idp}, starting from the existence conditions for the equilibria of~(\ref{eq: IDP}). The presented results are stated with respect to the saliency states $\upalpha=\Phi(z)$.
\begin{theorem}[Existence of retrievable memories]\label{thm:existence} Let $\upxi^{\uprho}$ be a memory as in Assumption~\ref{ass: W}. Then $x_{\uprho}=\upgamma_{\uprho}\upxi^{\uprho}$ is an equilibrium for~(\ref{eq: IDP}) for some unique $\upgamma_{\uprho}\in{\real}\setminus\{0\}$ if and only if $\upalpha_{\uprho}>1$. In this case, $\upgamma_{\uprho}$ satisfies
    \begin{align}\label{eq: eq_cond}
        \upalpha_\uprho=f(\upgamma_\uprho)=\tfrac{\upgamma_\uprho}{\uppsi(\upgamma_\uprho)}
    \end{align}
\end{theorem}
We now provide the main result on the local stability of the equilibria for the fast IDP Hopfield dynamics.
\begin{theorem}[Local stability of equilibria and memories]\label{thm: stb-cnd}
Consider the dynamics (\ref{eq: IDP}). Then
\begin{enumerate}
\item an equilibrium $x^{\star}\in{\real}^{N}$ is locally exponentially stable if and only if
    
    \begin{equation}
        -[D\Psi(x^{\star})]^{-1}+W(\upalpha)\prec{0}
    \end{equation}
    
    \item if $x_{\uprho}=\gamma_{\uprho}\xi^{\uprho}$ with $\upgamma_{\uprho}\in{\real}\setminus{\{0\}}$ is an equilibrium, then $x_{\uprho}$ is locally exponentially stable if and only if
    
    \begin{equation}
        {\uppsi'(\upgamma_{\uprho})}<\frac{1}{ \max_{\upmu=1,\dots,P}\:\upalpha_{\upmu}}
    \end{equation}
    \end{enumerate}
\end{theorem}
In particular, the study of the local stability condition allows to provide exact conditions on the points where the Jacobian becomes has eigenvalues with null real part, which will later be useful for the characterization of the memory manifold (see Section~\ref{sec: AP_Geo}). We will restrict ourselves to equilibrium conditions characterized by $\upgamma_\uprho>0$, as the case $\upgamma_\uprho<0$ is associated to the so called anti-memory $-\upxi^\uprho$ and is a symmetric artifact of classical Hopfield models. Notice in particular that by property~\ref{ass: act_f}(iv) the derivative $\uppsi'$ is locally invertible in $(0,+\infty)$, and we can define the following stability function.
\begin{definition}[Saliency stability function]\label{def: salienc}
    We define the saliency stability function $\upomega:\real_{>0}\to\real_{>0}$ as
    \begin{equation}
        \upomega(s)=f\ \circ\uppsi'^{-1}\left(\tfrac{1}{s}\right)= \frac{\uppsi'^{-1}\left(\tfrac{1}{s}\right)}{\uppsi\left(\uppsi'^{-1}\left(\tfrac{1}{s}\right)\right)}
    \end{equation}
\end{definition}
Before providing the key results on the eigenvalues of the Jacobian of the IDP Hopfield vector field, we establish the basic invertibility and monotonicity properties of the saliency stability function.
\begin{lemma}[Invertibility of the saliency stability function]\label{lem: sal_stab_prop}
    The saliency stability function $\upomega:\real_{>0}\to\real_{>0}$ is invertible, and in particular
    \begin{equation}
        \upomega^{-1}\left(r\right)=\frac{1}{\uppsi'\ \circ f^{-1} (r)},\qquad r>0.
    \end{equation}
    Furthermore, it holds that
    \begin{enumerate}
        \item\label{iit: ome} $\upomega(s)<s$;
        \item\label{iit: ome_inv} $\upomega^{-1}\left(r\right)>r$.
    \end{enumerate}
\end{lemma}
\begin{proof}
    The paragraph above Definition~\ref{def: salienc} establishes the invertibility of $\uppsi'$ in $(0,+\infty)$. Considering then the function $f(\upgamma)=\tfrac{\upgamma}{\uppsi(\upgamma)}$ we immediately observe that for any $\upgamma>0$ we have
    \begin{equation}
        f'(\upgamma)=\frac{\uppsi(\upgamma)-\upgamma\uppsi'(\upgamma)}{\uppsi'(\upgamma)^2}.
    \end{equation}
    From Assumption~\ref{ass: act_f}(iv) we have that $\uppsi'$ is monotonically decreasing on $(0,+\infty)$ and in particular we have
    \begin{align}\label{eq: psi_ord}
        \uppsi(\upgamma)&=\int_0^\upgamma \uppsi'(s)\ ds > \int_0^\upgamma \uppsi'(\upgamma)\ ds = \upgamma\uppsi'(\upgamma).
    \end{align}
    Consequently $f'(\upgamma)>0$ for all $\upgamma>0$ and in particular $f$ is invertible on $(0,+\infty)$. Taking $\upomega^{-1}(r)=1/(\uppsi'\ \circ f^{-1}(r))$ we observe that
    \begin{align}
        \upomega^{-1}\ \circ\upomega(s)&=\frac{1}{\uppsi'\circ f^{-1}\circ f\circ\uppsi'^{-1}\left(\tfrac{1}{s}\right)}\\
        &=\frac{1}{s^{-1}}=s
    \end{align}

    The proof of points~\ref{iit: ome}-\ref{iit: ome_inv} is straightforward, and relies on the fact that $\upgamma\uppsi'(\upgamma)-\uppsi(\upgamma)<0$ for all $\upgamma>0$, as proven in~(\ref{eq: psi_ord}).
    \begin{enumerate}
        \item $\upomega(s)<s$:\\
        We are interested in proving that
        \begin{equation}
            \upomega(s)-s=f\ \circ \uppsi'^{-1}\left(\tfrac{1}{s}\right)-s<0.
        \end{equation}
        Taking $\upgamma=\uppsi'^{-1}\left(\tfrac{1}{s}\right)$, we can express $s=\tfrac{1}{\uppsi'(\upgamma)}$, and therefore our condition reduces to
        \begin{equation}
            f(\upgamma)-\frac{1}{\uppsi'(\upgamma)}=\frac{\upgamma}{\uppsi(\upgamma)}-\frac{1}{\uppsi'(\upgamma)}<0.
        \end{equation}
        \item $\upomega^{-1}\left(r\right)>r$:\\
        We are interested in proving that
        \begin{equation}
            \tfrac{1}{\upomega^{-1}\left(r\right)}-\tfrac{1}{r}=\uppsi'\ \circ f^{-1}(r)-\tfrac{1}{r}<0.
        \end{equation}
        Taking $\upgamma = f^{-1}(r)$, we can express $r=f(\upgamma)$, and therefore our condition reduces to
        \begin{equation}
            \uppsi'(\upgamma)-\frac{1}{f(\upgamma)}=\uppsi'(\upgamma)-\frac{\uppsi(\upgamma)}{\upgamma}<0.
        \end{equation}
    \end{enumerate}
\end{proof}
The saliency stability function allows us to state an important proposition on the eigenvalues of the Jacobian of~\ref{eq: IDP} at the equilibrium points.
\begin{proposition}[Eigenvalues of the IDP Hopfield Jacobian]\label{prop: eig_jac}
    Let $\upalpha\in\real^{P}_{>0}$ and define $F(x)=-x+W(\upalpha)\Psi(x)$. 
    Let $\upalpha_\upnu>1$ such that for $\upgamma_\upnu=f^{-1}(\upalpha_\upnu)$ the equilibrium $x_{\upnu}=\upgamma_\upnu\upxi^\upnu$ exists. Then the Jacobian matrix
    \begin{equation}
        \frac{\partial F}{\partial x}(x_\upnu)=-\mI+W(\upalpha)\frac{\partial \Psi}{\partial x}(x_{\nu})
    \end{equation}
    has one eigenvalue $\uplambda_{\upmu}$ with $\Re\{\uplambda_\upmu\}=0$ iff
    \begin{equation}
        \exists\upmu\in\{1,\dots,P\}:\qquad\upalpha_\upnu=\upomega(\upalpha_\upmu).
    \end{equation}
\end{proposition}
\begin{proof}
    From~\cite[Theorems 2, SI]{BS-BG-FB-SZ:25idp} we have that
      \begin{equation}
        \frac{\partial F}{\partial x}(x_\upnu)=-\mI+W(\upalpha)\frac{\partial\Psi}{\partial x}(x_\upnu)
    \end{equation}
     This has the same spectrum of the matrix
     $$\frac{\partial\Psi}{\partial x}(x_\upnu)^{1/2}\frac{\partial F}{\partial x}(x_\upnu)\frac{\partial \Psi}{\partial x}(x_\upnu)^{-1/2}=
     \frac{\partial\Psi}{\partial x}(x_\upnu)^{1/2}\left(-\frac{\partial\Psi}{\partial x}(x_\upnu)^{-1}+W(\upalpha)\right)\frac{\partial\Psi}{\partial x}(x_\upnu)^{1/2}$$
        This matrix has the same inertia~\citep{HRA-JCR:85} of the matrix 
        $$DF(x_\upnu)=-\frac{\partial\Psi}{\partial x}(x_\upnu)^{-1}+W(\upalpha)$$
     that has spectrum
    \begin{equation}\label{eq: spec}
        \uprho\left(DF(x_\upnu)\right):=\Big{\{}\upalpha_1-\frac{1}{\uppsi'(\upgamma_\upnu)},\dots,\upalpha_P-\frac{1}{\uppsi'(\upgamma_\upnu)},\underbrace{-1,\dots,-1}_{N-P}\Big{\}},
    \end{equation}
    From this it is immediate to see that $\tfrac{\partial F}{\partial x}(x_\upnu)$ has null eigenvalues iff there exists $\upmu\in\{1,\dots,P\}$ such that
    \begin{equation}
        \uppsi'(\upgamma_\upnu)=\frac{1}{\upalpha_\upmu}.
    \end{equation}
    Exploiting the invertibility of $\uppsi'$ in $(0,+\infty)$, we get the equivalent condition
    \begin{equation}\label{eq: loc_stab}
        \upgamma_\upnu=\uppsi'^{-1}\left(\tfrac{1}{\upalpha_\upmu}\right)\quad \exists \upmu\in\{1,\dots,P\}.
    \end{equation}
    Since $\upalpha_\upnu>1$, then exploiting the uniqueness of $\upgamma_\upnu>0$ in the equilibrium condition in~(\ref{eq: eq_cond}) we have that the zero eigenvalue condition in~(\ref{eq: loc_stab}) coincides with
    \begin{align}\label{eq: zero_eig}
        \upalpha_\upnu&=f(\upgamma_\upnu)\nonumber\\
        &=f\ \circ\uppsi'^{-1}\left(\tfrac{1}{\upalpha_\upmu}\right)=\upomega(\upalpha_\upmu)\quad \exists \upmu\in\{1,\dots,P\}.
    \end{align}
\end{proof}
In particular, notice that from Lemma~\ref{lem: sal_stab_prop} for all $\upnu=1,\dots,P$ it holds that $\upalpha_\upnu>\upomega(\upalpha_\upnu)$, so that whenever at least a memory exists, at least one is also locally exponentially stable. Notice in addition that the number of eigenvalues with positive and negative real part is obtained by counting the instances where the condition $\upalpha_\upnu \lessgtr \upomega(\upalpha_\upmu)$ is satisfied for $\upmu=1,\dots,P$. 

\section{A modern reformulation of the sequential IDP Hopfield model}
\label{ssec:modern_IDP}

The IDP Hopfield model admits a modern Hopfield \citep{KD-HJJ:20} interpretation in which retrieval is decomposed into three interacting processes: a memory layer that reads out the similarity between the current feature state and the stored memories, a saliency layer that modulates this readout, and a feature layer that reconstructs the retrieved representation~\citep{BS-BG-FB-SZ:25idp}. In the original IDP model, the saliency layer is driven by the external input. Here we show that the sequential IDP Hopfield model~(\ref{eq: ttreas}) admits the same interpretation, with the saliency layer driven instead by internal graph-structured feedback.

We introduce an auxiliary memory-index variable $y\in\real^P$ and consider
$$
\begin{cases}
\uptau_x \dot x = -x + M\diag(\Phi(z))y,\\
\uptau_y \dot y = -y + M^\top\Psi(x),\\
\uptau_z \dot z = -z + \chi(y).
\end{cases}
$$
The variable $y$ is the memory-layer activity: its components encode the alignment between the current activated feature state and the stored memories. The map $\chi:\real^P\to\real^P$ is a slow feedback map prescribing the target saliency profile toward which $z$ relaxes.

Under the timescale separation $\uptau_y\ll\uptau_x\ll\uptau_z$, the memory layer rapidly satisfies
$$
y\simeq M^\top\Psi(x).
$$
Substituting this quasi-steady relation into the feature dynamics gives
$$
\uptau_x\dot x=-x+M\diag(\Phi(z))M^\top\Psi(x),
$$
which is exactly the fast equation of the sequential IDP Hopfield model~(\ref{eq: ttreas}).

The slow equation in~(\ref{eq: ttreas}) is recovered by choosing
$$
\chi(y)=\frac{\kappa}{\sqrt N}Ay.
$$
Indeed, replacing $y$ by its fast quasi-steady value yields
$$
\uptau_z\dot z=-z+\frac{\kappa}{\sqrt N}AM^\top\Psi(x),
$$
which coincides with the slow feedback dynamics of~(\ref{eq: ttreas}).

This reformulation clarifies the relation with the modern IDP interpretation. In the original IDP model, saliencies are prescribed by an external input decomposition. In the sequential IDP model, saliencies are generated autonomously by the retrieved memory through the transition graph. 

\section{Memory manifolds: pouring the concrete}\label{sec: AP_Geo}
In the classic slow-fast characterization from geometric singular perturbation theory (GSPT) presented in~\cite{FN:79}, the timescale separation between coupled dynamics is extremized in order to derive geometric information about the loci in state-space where the system evolves. In the case of the IDP Hopfield model, the coupled system can be written as
\begin{equation}\label{eq: gspt-HOP_SI}
\begin{cases}
\varepsilon \dot x = -x+M\,\diag(\upphi(z))\,M^{\top}\Psi(x), \quad\:\:\:\: x(0)\in\mathbb{R}^N,\\
\dot z = -z + \dfrac{\kappa}{\sqrt{N}}\, A M^{\top}\Psi(x), \qquad\qquad\quad\ \ z(0)\in\mathbb{R}^P.
\end{cases}
\end{equation}
and taking the limit $\varepsilon\to 0^+$, the problem simplifies to studying how the dynamics
\begin{equation}
    \dot z = -z+\frac{\kappa}{\sqrt{N}}AM^\top\Psi(x^\star(z))
\end{equation}
evolve under the solution constraint, parametrized as a function of the slow variable $z$, $\vectorzeros[N]=-x^\star(z)+M\diag(\Phi(z))M^\top\Psi(x^\star(z))$. We start by studying the solutions of the algebraic constraint, which characterize the critical set
\begin{equation}
    \mathcal{S}_0:=\{(x,z)\in\real^N\times\real^P:\: x=M\diag(\Phi(z))M^\top\Psi(x)\}
\end{equation}
\subsection{Memory branches and memory manifolds}
Exploiting Theorem~\ref{thm:existence} for the existence of equilibria in the IDP Hopfield model, we proceed to define the memory branches, which are disjoint sets in $\mathcal{S}_0$ associated each to a different memory.
\begin{definition}\label{def: memory_branch}
    Let $\upphi(z_\upnu)>1$. Then the $\upnu^{\text{{th}}}$ memory branch is the $P$-dimensional set $\mathcal{N}_\upnu\subseteq \mathcal{S}_0$ defined as
\begin{equation}\label{def: memory_branch}
\mathcal{N}_{\upnu}
=\left\{(x,z)\in\mathbb{R}^N\times\mathbb{R}^P :\ \exists\upgamma_\upnu\neq 0,\:  
x=\upgamma_\upnu \upxi^\upnu,\ 
\upphi(z_\upnu)=f(\upgamma_\upnu)
\right\},
\end{equation}
\end{definition}
Notice that the memory branch $\mathcal{N}_\upnu$ is not defined for $\upphi(z_\upnu)<1$, as the equilibria associated to the memory $\upxi^\upnu$ does not exists. The memory manifold is then defined as a $P$-dimensional manifold comprising the points in $\mathcal{N}_\upnu$ where normal hyperbolicity holds.
\begin{definition}\label{def: memory_man}
    Let $F(x,z)=-x+W(\Phi(z))\Psi(x)$ and assume $\upphi(z_\upnu)>1$. Then the $\upnu^{\text{{th}}}$ memory manifold $\mathcal{M}_\upnu$ is the set of points $(x,z)\in\mathcal{N}_\upnu$ such that
    \begin{equation}
        \nexists\uplambda\in\uprho\left(\frac{\partial F}{\partial x}(x,z)\right):\quad \Re(\lambda)=0.
    \end{equation}
\end{definition}
Of particular relevance are two disjoint regions of the manifold $\mathcal{M}_\upnu$, which play qualitatively distinct roles in shaping the system’s long-term behavior. Rather than pursuing formal definitions, we give an intuitive description aimed at conveying their dynamical function:
\begin{itemize}
    \item \emph{Attracting manifold} $\mathcal{M}_\upnu^a$: this region consists of points that act as local sinks for the dynamics. Trajectories initialized in a neighborhood of these points in the ambient space are drawn toward the manifold and remain confined to it over time. Geometrically, $\mathcal{M}_\upnu^a$ behaves as a stabilizing surface that captures and retains nearby trajectories.
    \item \emph{Saddle manifold} $\mathcal{M}_\upnu^s$: this region contains points that are simultaneously attracting and repelling, depending on direction. Trajectories may approach the manifold transiently, but the presence of unstable directions eventually pushes them away. As a result, $\mathcal{M}_\upnu^s$ acts as a transient conduit, enabling trajectories to leave one attracting region and move toward another.
\end{itemize}
One of the advantages of relying on the IDP Hopfield model to capture sequentiality and memory transition is that there already exists a complete theory on the stability of system's equilibria. We exploit exactly this theory to characterize the attracting and saddle memory manifolds.
\begin{lemma}[Attracting and saddle manifolds]
    Let $\upomega:\real_{>0}\to\real_{>0}$ be the saliency stability function in Definition~\ref{def: salienc}. Then the point $(x,z)\in\mathcal{M}_\upnu$ on the memory manifold  is
    \begin{itemize}
        \item attracting if $\upphi(z_\upnu)>\upomega(\upphi(z_\upmu))$ for all $\upmu\in\{1,\dots,P\}$;
        \item saddle type if there exists at least one $\upmu\in\{1,\dots,P\}/\{\upnu\}$ such that $\upphi(z_\upnu)<\upomega(\upphi(z_\upmu))$.
    \end{itemize}
\end{lemma}
\begin{proof}
    The manifold $\mathcal{M}_\upnu$ is attracting at $(x,z)$ if for all
    \begin{equation}
        \uplambda\in\uprho\left(\frac{\partial F}{\partial x}(x,z)\right)\quad \Re(\uplambda)<0
    \end{equation}
    and a saddle if there exists at least an eigenvalue $\uplambda$ such that $\Re(\uplambda)>0$. From Proposition~\ref{prop: eig_jac}, we have that the eigenvalues of the Jacobian $\tfrac{\partial F}{\partial x}$ at $(x,z)\in\mathcal{M}_\upnu$ have the same sign of $\{\upphi(z_\upmu)-\tfrac{1}{\uppsi'(\upgamma_\upnu)}\}_{\upmu=1}^P$ with $\upgamma_\upnu=f^{-1}(\upphi(z_\upnu))$. Thus, $\mathcal{M}_\upnu$ is attracting at $(x,z)$ iff for all $\upmu=1,\dots,P$
    \begin{equation}
        \upphi(z_\upmu)-\frac{1}{\uppsi'(\upgamma_\upnu)}<0\quad\iff\quad \upphi(z_\upnu)>\upomega(\upphi(z_\upmu)).
    \end{equation}
    Conversely, it is a saddle if there exists $\upmu\in\{1,\dots,P\}/\{\upnu\}$ such that
    \begin{equation}
        \upphi(z_\upmu)-\frac{1}{\uppsi'(z_\upnu)}>0\quad\iff\quad \upphi(z_\upnu)<\upomega(\upphi(z_\upmu)).
    \end{equation}
\end{proof}
The phenomena known as retrieval and fixation to the memory $\upxi^\upnu$ occur for points $(x,z)\in \mathcal{M}_\upnu$ such that the memory manifold is attracting, i.e., the respective saliency weight $\upphi(z_\upnu)$ is greater than the saliency stability function computed at any $\upphi(z_\upmu)$. Escape direction emerge instead when the dynamics evolve towards a saddle region of the memory manifold. In general, we have that the attracting manifold for the memory $\upxi^\upnu$ is the subset $\mathcal{M}^a_\upnu:=\{(x,z)\in\mathcal{M}_\upnu:\ \upphi(z_\upnu)>\upomega(\upphi(z_\upmu))\quad \forall\upmu=1,\dots,P\}$. Conversely, the saddle manifold is $\mathcal{M}^s_\upnu:=\{(x,z)\in\mathcal{M}_\upnu:\ \exists\upmu=1,\dots,P\quad \upphi(z_\upnu)<\upomega(\upphi(z_\upmu))\}$.

\subsection{The origin manifold}
In our description of the geometric sets where the slow-fast activity of the system is constrained, we considered all the equilibria solutions of the fast IDP Hopfield dynamics that are associated to one of the memories. However, there is another solution that is independent from any of the memory vectors: the origin. 

\begin{definition}[Origin manifold]
    The origin manifold is defined as the set
    \begin{equation}
        \mathcal{M}_0:=\Big\{(\vectorzeros[N],z)\in\real^N\times\real^P:\quad\nexists\lambda\in\uprho\left(\frac{\partial F}{\partial x}(\vectorzeros[N],z)\right),\ \Re\{\uplambda\}=0\Big\}.
    \end{equation}
    In particular, for any point in the set $\mathcal{M}_0$ we have that
    \begin{equation}
        \frac{\partial F}{\partial x}(\vectorzeros[N],z)=-\mI+M\diag(\Phi(z))M^\top
    \end{equation}
\end{definition}
Notice that the structure of the Jacobian $\tfrac{\partial F}{\partial x}(\vectorzeros[N],z)$ allows for a clean characterization of the attracting and saddle regions. Noticing that the spectrum of Jacobian is
\begin{equation}
    \uprho\left(\tfrac{\partial F}{\partial x}(\vectorzeros[N],z)\right):=\{-1+\upphi(z_1),\dots,-1+\upphi(z_P),\underbrace{-1,\dots,-1}_{N-P}\},
\end{equation}
we can characterize the not normally hyperbolic points as $(\vectorzeros[N],z)$ such that there exists $\upmu=1,\dots,P$ for which $\upphi(z_\upmu)=1$. Similarly, we can characterize the:
\begin{enumerate}
    \item attracting origin manifold: $\mathcal{M}_0^a:=\{(\vectorzeros[N],z):\quad \|\Phi(z)\|_\infty<1\}$;
    \item saddle origin manifold: $\mathcal{M}_0^s:=\mathcal{M}_0\setminus\mathcal{M}_0^a$.
\end{enumerate}
Of particular relevance is the attracting origin manifold $\mathcal{M}_0^a$, since it is a sink for the slow-fast dynamics, and once trajectories enter this set they cannot escape it. Thereby, from $\mathcal{M}_0^a$ the activity inevitably collapses to the origin $(\vectorzeros[N],\vectorzeros[P])\in\real^N\times\real^P$, as formalized in the following Lemma.
\begin{lemma}[Collapse to the origin]
    Let $(x_0,z_0)\in\mathcal{M}_0^a$. Then $(x(t),z(t))\to(\vectorzeros[N],\vectorzeros[P])$ exponentially fast.
\end{lemma}
\begin{proof}
        For $(x_0,z_0)\in\mathcal{M}_0^a$ the slow dynamics reduce to
        \begin{align}
            \dot z &=-z+\frac{\kappa}{\sqrt{N}}AM^\top\Psi(\vectorzeros[N])\nonumber\\
            &=-z.
        \end{align}
        Therefore, the slow dynamics are completely dissipative and $z(t)=z_0 e^{-t}$.
\end{proof}
It is easy to observe that, in particular, $\mathcal{M}_0^a$ is also forward-invariant for the reduced dynamics.


\section{Slow saliency dynamics: under the hood}
\label{app:slow-reas}

We provide the technical details for Sec.~\ref{sec:slow-reasoning}. We first prove Theorem~\ref{thm:escape-time}, and then analyze the explicit scalar map introduced in the example.

\subsection{Proof of Theorem~\ref{thm:escape-time}}

We start with two auxiliary results. The first one is a corollary of Lemma~\ref{lem: sal_stab_prop}.

\begin{corollary}\label{lemma1}
For every $z$ such that $\upphi(z)>1$, we have
\[
\uppsi'\bigl(f^{-1}(\upphi(z))\bigr)<\frac{1}{\upphi(z)}.
\]
\end{corollary}

\begin{proof}
From Lemma~\ref{lem: sal_stab_prop} we know that
\begin{equation}
    \upomega^{-1}\left(r\right)>r.
\end{equation}
Taking $r=\upphi(z)$ and the definition $\upomega^{-1}\left(r\right)=1/(\uppsi'\ \circ f^{-1}(r))$ we conclude.
\end{proof}

\begin{figure}[!h]
\begin{center}
\includegraphics[width=0.8\linewidth]{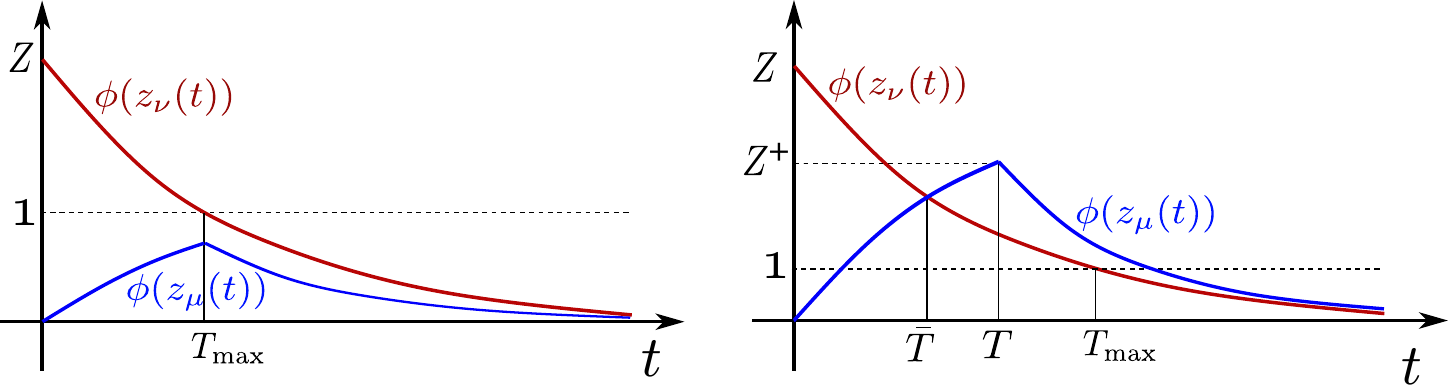}
\end{center}
\caption{Graphical illustration of the two possible outcomes of the source-successor saliency dynamics. Left: no transition occurs and the dynamics decay to inactivity. Right: the successor saliency catches up with the source saliency, triggering a transition between memories.}
\label{fig:zampi}
\end{figure}

In the two-memory reduction of Sec.~\ref{sec:slow-reasoning}, the source and successor saliencies satisfy
$$
z_\upnu(t)=Ze^{-t},\qquad \dot z_\upmu=-z_\upmu+\kappa g(Ze^{-t}),\qquad g:=\uppsi\circ f^{-1}\circ\upphi .
$$
Hence
\begin{equation}\label{eq:app-zmu-solution}
z_\upmu(t)=\kappa\int_0^t e^{-(t-\tau)} g(Ze^{-\tau})\,d\tau.
\end{equation}
Let $T_{\max}$ be defined by $\upphi(Ze^{-T_{\max}})=1$. The next lemma shows that, once the successor saliency catches up with the source saliency, the source branch must lose stability before the source memory disappears (see also Fig.~\ref{fig:zampi}, right plot).

\begin{lemma}\label{lemma2}
Assume that there exists $\bar T\in[0,T_{\max})$ such that
$z_\upmu(\bar T)=z_\upnu(\bar T)$. Then there exists
$T\in[\bar T,T_{\max})$ such that
\begin{equation}\label{eq:lemma2}
    \upomega^{-1}(\upphi(z_\upmu(T))>\upphi(z_\upmu(T))=\upomega^{-1}(\upphi(z_\upnu(T)).
\end{equation}
\end{lemma}

\begin{proof}
Observe first that the inequality on the left is always true by Corollary~\ref{lemma1}. We need to prove that the equality holds true for some $T\in[\bar T,T_{\max})$. 
To this aim we will prove that
\begin{align}
    &\upphi(z_\upmu(t))<\upomega^{-1}(\upphi(z_\upnu(t))\quad\text{for $t=\bar T$}\label{eq:1}\\
    &\upphi(z_\upmu(t))>\upomega^{-1}(\upphi(z_\upnu(t))\quad\text{for $t=T_{\max}^-$}\label{eq:2}
\end{align}
which would prove that the equality should hold between $\bar T$ and $T_{\max}$.

Note that, again by Corollary \ref{lemma1}, we have 
$$\upphi(z_\upmu(\bar T))=\upphi(z_\upnu(\bar T))<\upomega^{-1}(\upphi(z_\upnu(\bar T))$$
We want to prove now (\ref{eq:2}). 
To this aim let $\delta(t):=z_\upmu(t)-z_\upnu(t)$ and observe  that $\delta(t)$ satisfies the same differential equation satisfied by $z_\upmu(t)$ and hence we can argue that $\delta(t)$ is strictly increasing and $\delta(\bar T)=0$. This implies that $\delta(T_{\max}^-)>0$ and hence
$$\frac{1}{\upphi(z_\upmu(T_{\max}^-))}=\frac{1}{\upphi(z_\upnu(T_{\max}^-)+\delta(T_{\max}^-))}<\frac{1}{\upphi(z_\upnu(T_{\max}^-))}=1$$
On the other hand observe that $\upphi(z_\upnu(T_{\max}^-))=1$ so that $ f^{-1}(\upphi(z_\upnu(T_{\max}^-)))=0$ and hence
$$\upomega^{-1}(\upphi(z_\upnu(T_{\max}^-))=1$$
where we used the assumption $\uppsi'(0)=1$. 

This yields the thesis.
\end{proof}

We now prove Theorem~\ref{thm:escape-time}.

\begin{proof}[Proof of Theorem~\ref{thm:escape-time}]
By Lemma~\ref{lemma2} and the IDP Hopfield stability condition in Theorem~\ref{thm: stb-cnd}, if the successor saliency catches up with the source saliency before $T_{\max}$, then there exists a time in $[\bar T,T_{\max})$ at which the source branch loses stability. The first such time, $T_{\mathrm{esc}}$, satisfies the identity in~(\ref{eq:lemma2}), namely
\[
\upphi(z_\upmu(T_{\mathrm{esc}}))=\upomega^{-1}(\upphi(z_\upnu(T_{\mathrm{esc}})).
\]
Equivalently,
\[
z_\upmu(T_{\mathrm{esc}})
=
h\bigl(z_\upnu(T_{\mathrm{esc}})\bigr),
\qquad
h(z):=
\upphi^{-1}\!\left(
\upomega^{-1}(\upphi(z))
\right).
\]
Using $z_\upnu(t)=Ze^{-t}$ and (\ref{eq:app-zmu-solution}), and evaluating the previous identity at $t=T_{\mathrm{esc}}$ yields
\[
\kappa\int_0^{T_{\mathrm{esc}}}
e^{-(T_{\mathrm{esc}}-\tau)}
g(Ze^{-\tau})\,d\tau
=
h\bigl(Ze^{-T_{\mathrm{esc}}}\bigr).
\]
Thus $T_{\mathrm{esc}}$ is the smallest admissible solution of the previous equation, and the updated saliency is
\[
Z^+:=z_\upmu(T_{\mathrm{esc}})
=
h\bigl(Ze^{-T_{\mathrm{esc}}}\bigr).
\]
\end{proof}

\subsection{Extended derivations for the illustrative example}

We give the details for the scalar iteration associated with
$$
\uppsi(x)=\frac{x}{1+|x|},\qquad\upphi(z)=|z|.
$$
In this case it can be proved that $f(x)=\frac{x}{\uppsi(x)}=x+1$ for all $x\ge 0$ so that, for all $x\ge 1$ we have that $f^{-1}(x)=x-1$. Moreover for $x\ge 0$ we have $\uppsi'(x)=\frac{1}{(x+1)^2}$. Thus, for $z\ge1$,
$$
g(z)=1-z^{-1}\qquad h(z)=z^2.
$$

Consequently, for $Z>1$ and $X\in[1,Z)$,
$$
\int_1^X g\!\left(\frac{Z}{s}\right)ds=\int_1^X\left(1-\frac{s}{Z}\right)ds=(X-1)\left(1-\frac{X+1}{2Z}\right).
$$
Substituting this expression into $R(Z,X)=0$ yields
$$
\frac{\kappa}{X}(X-1)\left(1-\frac{X+1}{2Z}\right)=\frac{Z^2}{X^2},
$$
so that the saliency update map in (\ref{eq:saliency-map}) is determined by
\begin{equation}\label{eq:app-example-map}
\begin{cases}
G(Z,X)=\frac{1}{\kappa},\\[1mm]
Z^+=\frac{Z^2}{X^2},
\end{cases} \qquad X\in[1,Z),
\end{equation}
where
\begin{equation}\label{eq:functionG}
G(Z,X):=-\frac{X(X-1)(X+1-2Z)}{2Z^3}.
\end{equation}
For a given $Z$, the admissible value of $X$ is the smallest solution of $G(Z,X)={1}/{\kappa}$ in $[1,Z)$, if such a solution exists.

\begin{figure}[!h]
\begin{center}
\includegraphics[width=0.5\linewidth]{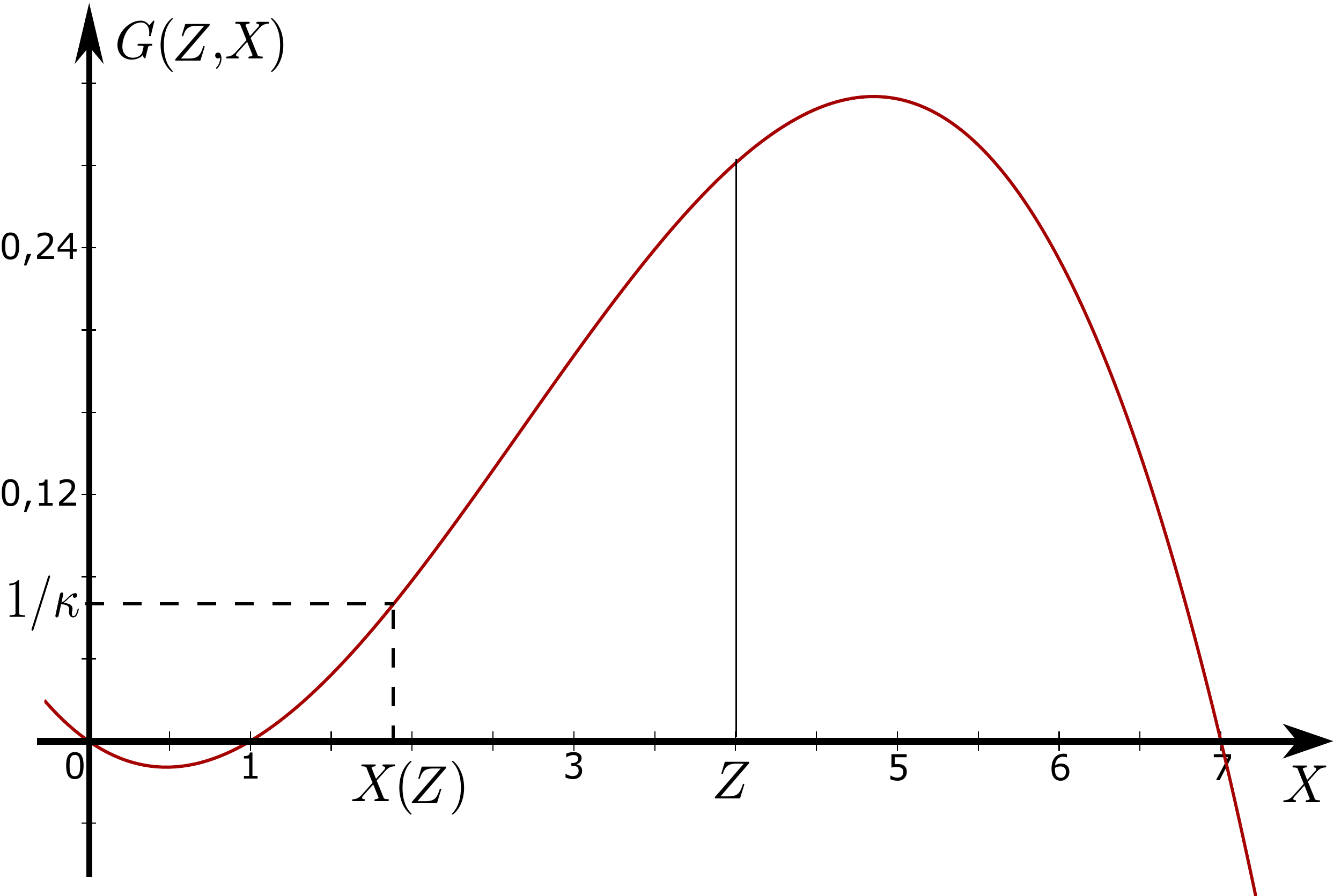}
\end{center}
\caption{Plot of the function $G(Z,X)$ in (\ref{eq:functionG}) for $\kappa=15$ and $Z=4$.}
\label{fig:zampi2}
\end{figure}

\paragraph{Existence of the iteration.}
We present a condition on $Z$ ensuring that the iteration in (\ref{eq:app-example-map}) exists, namely that there exists a solution $X(Z)\in[1,Z)$ of the equation $G(Z,X)=1/\kappa$ (see also Fig.~\ref{fig:zampi2}). Observe that $G(Z,1)=0$ and that $G(Z,Z)=\frac{1}{2}\left(\frac{Z-1}{Z}\right)^2$. This implies that, by taking $Z$ such that
\begin{equation}\label{eq:existence}
    \frac{1}{\kappa}<\frac{1}{2}\left(\frac{Z-1}{Z}\right)^2\quad\Longleftrightarrow\quad Z>\frac{1}{1-\sqrt{2/\kappa}}
\end{equation}
a solution $X(Z)\in[1,Z)$ must exist. Such solution is unique. Indeed, since
\begin{equation}\label{eq:partderivG}
\frac{\partial G}{\partial X}=-\frac{3X^2-4ZX-1+2Z}{2Z^3}
\end{equation}
the sign of the this derivative coincides with the sign of the function $q_Z(X):=-(3X^2-4ZX-1+2Z)$ that is a concave parabola. Since $q_Z(1)=2(Z-1)$ and $q_Z(Z)=(Z-1)^2$ that are both positive, we have that $q_Z(X)$ is positive in the interval $[1,Z)$. Hence $G(Z,X)$ is strictly increasing in that interval, so whenever (\ref{eq:existence}) holds there is a unique solution $X(Z)\in[1,Z)$ and $Z^+=Z^2/X(Z)^2$ is uniquely defined.

\begin{figure}[!h]
\begin{center}
\includegraphics[width=0.5\linewidth]{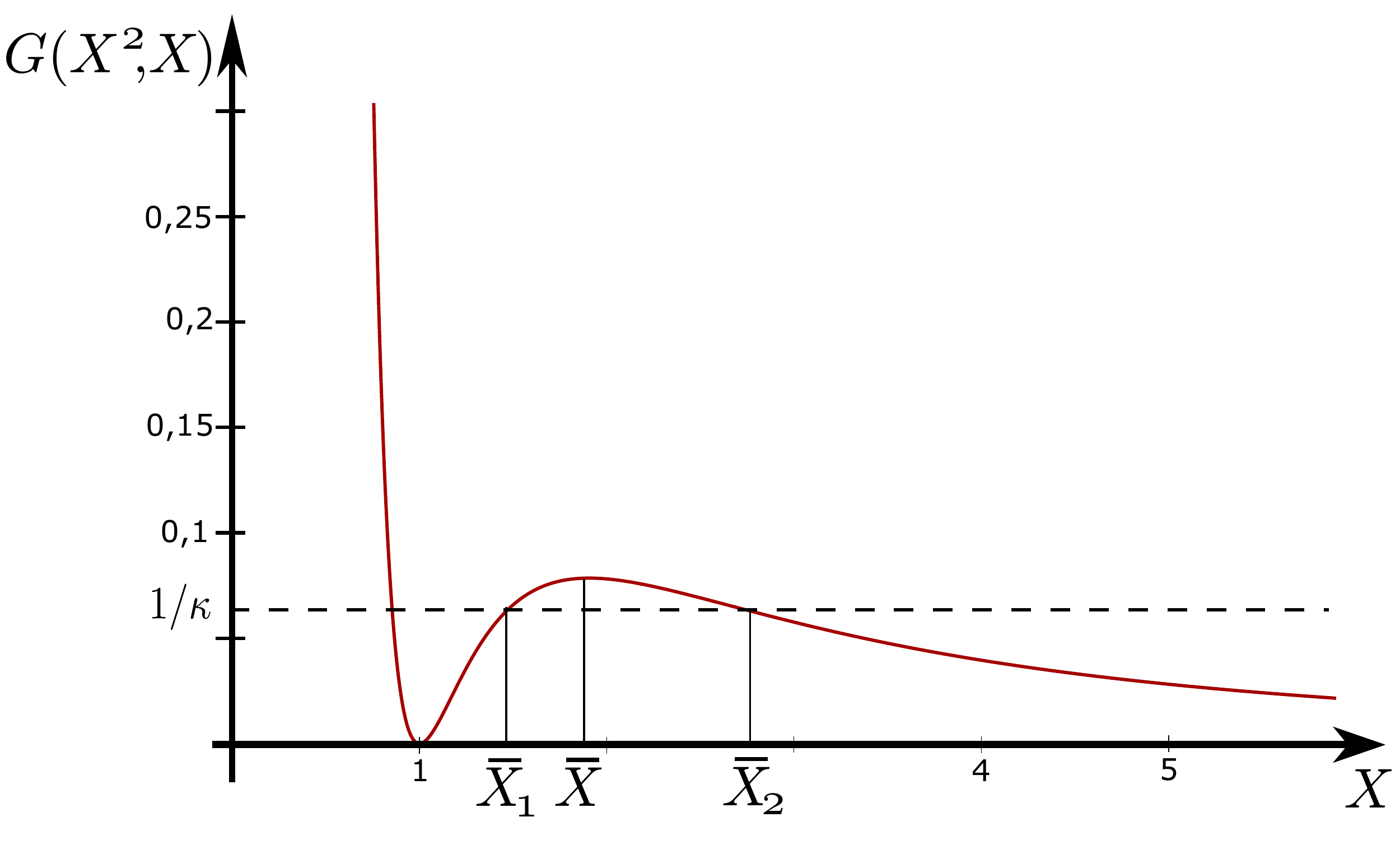}
\end{center}
\caption{Plot of the function $G(X^2,X)$ in (\ref{eq:fixpointsX}) and its intersections with the line $1/\kappa$.}
\label{fig:zampi1}
\end{figure}

\paragraph{Fixed points of the iteration.}
Fixed points satisfy $Z^+=Z$. Since $Z^+=Z^2/X^2$, this gives $Z=X^2$. Substituting into $G(Z,X)=1/\kappa$ yields (see also Fig.~\ref{fig:zampi1})
\begin{equation}\label{eq:fixpointsX}
\frac{1}{\kappa}=G(X^2,X)=\frac{(X-1)^2(2X+1)}{2X^5}
\end{equation}
Observe that
\begin{equation}\label{eq:derivG}
    \frac{d G(X^2,X)}{d X}=-\frac{(X-1)}{2X^6}(4X^2-5X-5)
\end{equation}
If we let $\bar X=\frac{5+\sqrt{105}}{8}\simeq1.9059$, then we have that $G(X^2,X)$ is strictly increasing for $1\le X<\bar X$ the function, and strictly decreasing for $X>\bar X$, so that $G(X^2,X)$ reaches its maximum in $\bar X$. Since  $G(\bar X^2,\bar X)\simeq0.0785$, we can argue that we have fixed points only when
\begin{equation*}
    \kappa\ge\kappa_{\mathrm{crit}}:=\frac{1}{G(\bar X^2,\bar X)}\simeq 12.74
\end{equation*}
For $\kappa>\kappa_{\mathrm{crit}}$, there are exactly two solutions, $\bar X_1$ and $\bar X_2$, of (\ref{eq:fixpointsX}) which satisfy $1<\bar X_1<\bar X<\bar X_2$. Such solutions in turn yield two fixed points, $\bar Z_1=\bar X_1^2$ and $\bar Z_2=\bar X_2^2$ which satisfy $1<\bar Z_1<\bar Z<\bar Z_2$ with $\bar Z=\bar X^2=3.6323$.

\paragraph{Monotonicity of the iteration.} We will now show that the iteration $Z^+=\frac{Z^2}{X(Z)^2}$ is strictly increasing in $Z$, which proves that the sequence of iterates, denoted by $Z_0,Z_1,Z_2,\ldots$, is monotonic and hence it converges to a limit point $Z_\infty$ .
First notice that
$$\frac{\partial Z^+}{\partial Z}=\frac{2Z}{X^2}-\frac{2Z^2}{X^3}\frac{\partial X}{\partial Z}=\frac{2Z}{X^2}+\frac{2Z^2}{X^3}\frac{\partial G/\partial Z}{\partial G/\partial X}=\frac{2Z}{X^3}\frac{1}{\partial G/\partial X}\left(X\frac{\partial G}{\partial X}+Z\frac{\partial G}{\partial Z}\right)
$$
Using
$$
\frac{\partial G}{\partial Z}=\frac{X(X-1)(3X-4Z+3)}{2Z^4},
$$
we obtain
$$
X\frac{\partial G}{\partial X}+Z\frac{\partial G}{\partial Z}=\frac{X(Z-1)}{Z^3},
$$
and hence
$$
\frac{\partial Z^+}{\partial Z}=\frac{2(Z-1)}{X^2Z^2}\frac{1}{\partial G/\partial X}.
$$
That shows that the sign of $\partial Z^+/\partial Z$ only depends on the sign of $\partial G/\partial X$. But, if the existence condition (\ref{eq:existence}) holds, the derivative $\partial G/\partial X$ in (\ref{eq:partderivG}) evaluated at $X(Z)\in[1,Z)$ must be positive.
Thus, we conclude that iteration $Z^+=\frac{Z^2}{X(Z)^2}$ is strictly increasing if $Z>{1}/({1-\sqrt{2/\kappa}})$.

\paragraph{Convergence of the iteration.} We show finally that the fixed point $\bar Z_1$ is repelling (or unstable) and that $\bar Z_2$ is attractive (or stable). To this aim observe that, at the fixed points, where $Z=X^2$, we have
$$\frac{\partial Z^+}{\partial Z}=\frac{4(X+1)}{4X^2-X-1}\quad\Longrightarrow\quad  \frac{\partial Z^+}{\partial Z}-1=-\frac{4X^2-5X-5}{4X^2-X-1}.$$
Then $\frac{\partial Z^+}{\partial Z}>0$. Moreover, comparing the expression of $\frac{\partial Z^+}{\partial Z}-1$ with (\ref{eq:derivG}), we see that, since $4X^2-X-1>0$, then $\frac{\partial Z^+}{\partial Z}-1$ has the same sign of $\frac{d G(X^2,X)}{d X}$ showing in this way that $\frac{\partial Z^+}{\partial Z}>1$ in $X=\bar X_1$ and $0<\frac{\partial Z^+}{\partial Z}<1$ in $X=\bar X_2$. This proves that the fixed point $\bar Z_1$ is repelling while the fixed point $\bar Z_2$ is attractive. 

The previous analysis, together with the monotonicity of the iteration, shows that amplitudes above the unstable threshold $\bar{Z}_1$ converge to the attracting fixed point $\bar{Z}_2$, which corresponds to self-sustained sequential retrieval. In contrast, amplitudes below $\bar{Z}_1$ decay until subsequent transitions can no longer be triggered, ultimately leading the dynamics to collapse into the inactive state. This is illustrated in Fig.~\ref{fig: z_strength} in the main text. 

\section{Additional results for Tanh activation}
\label{app:tanh}

\begin{figure}[h]
    \centering
    \subfloat{\includegraphics[width=0.35\linewidth]{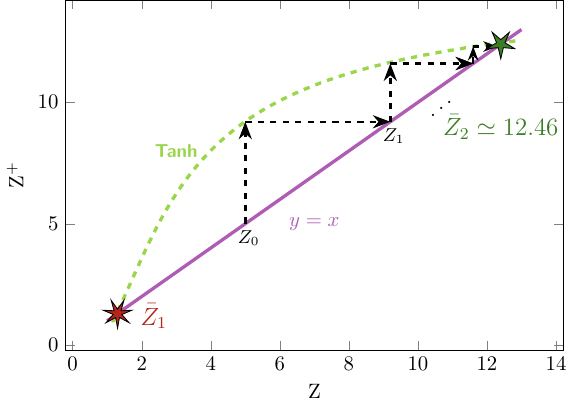}}\hspace{1cm}
    \subfloat{\includegraphics[width=0.35\linewidth]{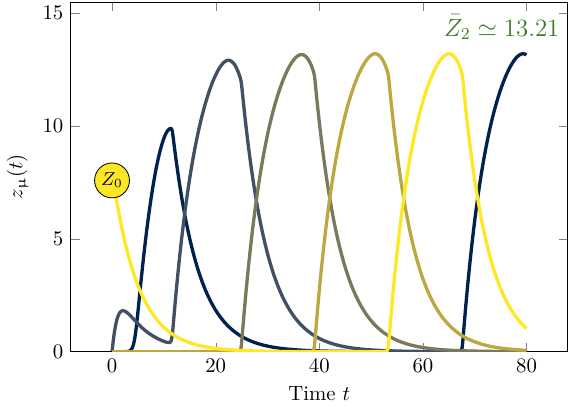}}
    \caption{Discrete iteration map $Z^+=h(Z/X)$ for the $\uppsi=\tanh$ activation function and associated $z$-components for the case of $\kappa=15$.}
    \label{fig:tanh-app}
\end{figure}

We complement the analysis of Sec.~\ref{sec:slow-reasoning} by considering the Tanh activation $\psi(x)=\tanh(x)$, while retaining $\upphi(z)=|z|$. Unlike SmoothF, the corresponding saliency map and escape-time predictions do not admit simple closed-form expressions, but can be computed numerically from the implicit equations derived in Sec.~\ref{sec:slow-reasoning}. Fig.~\ref{fig:tanh-app} illustrates the resulting discrete map and representative saliency dynamics for $\kappa=15$.

\begin{figure}[h]
    \centering
    \includegraphics[width=\linewidth]{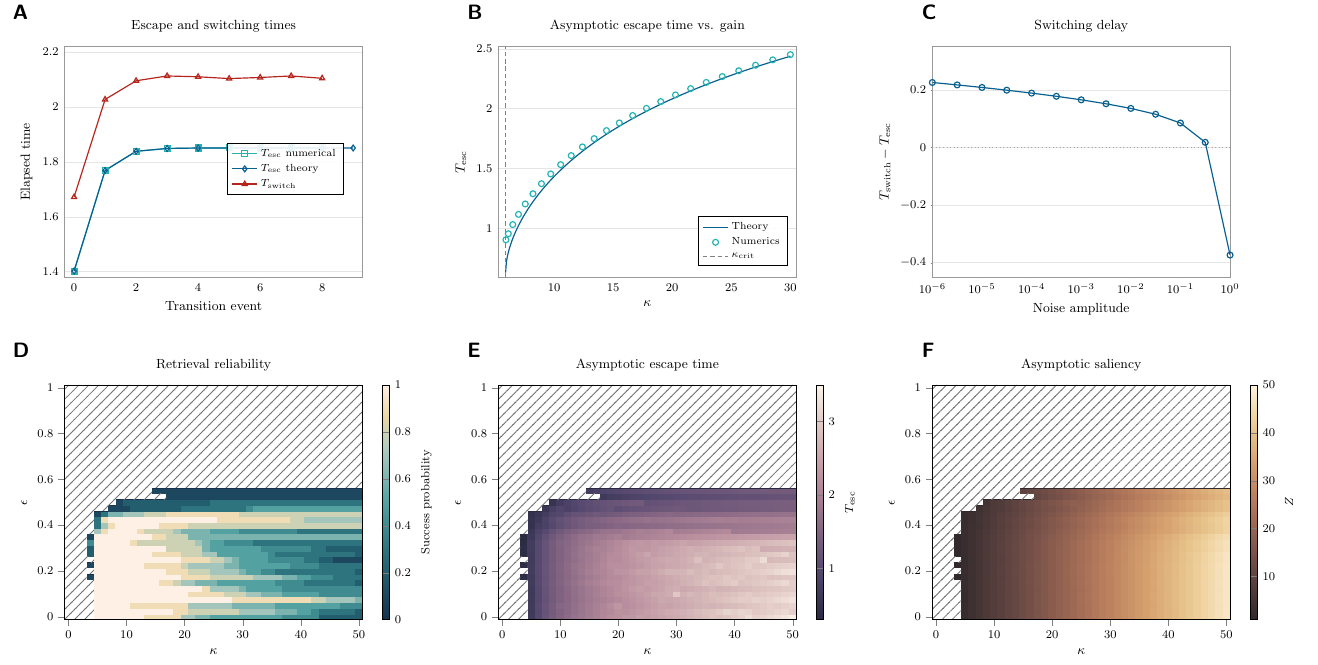}
    \caption{
    Numerical validation and robustness analysis for Tanh activation.
    (A) Theoretical and numerical escape times and observed switching times across consecutive transitions.
    (B) Asymptotic escape time versus $\kappa$: theory and numerical estimates.
    (C) Mean switching delay versus Gaussian noise amplitude.
    (D) Probability of correct transitions in the $(\kappa,\epsilon)$ plane.
    (E, F) Asymptotic escape time and saliency amplitude, respectively.
    All experiments use the same numerical configuration as Fig.~\ref{fig:corr_main}. Hatched regions in (D-F) indicate no successful transitions.
    }
    \label{fig:tanh-corr}
\end{figure}

We also repeat the numerical experiments of Fig.~\ref{fig:corr_main} using Tanh activation (Fig.~\ref{fig:tanh-corr}). For orthogonal memories, theoretical and numerical escape times remain in good agreement (A, B), while increasing the noise amplitude reduces the switching delay and eventually leads to premature switching (C). For non-orthogonal memories, retrieval reliability decreases with memory bias and exhibits a non-monotonic dependence on the gain $\kappa$, with excessively large gains reducing the probability of successful transitions (D). Conditional on successful transitions, both the asymptotic escape time and saliency generally increase with $\kappa$ (E, F). These results support the applicability of the theoretical framework to different activation functions, while highlighting activation-dependent differences in retrieval robustness.

\section{Numerical investigation of sequential retrieval capacity}
\label{app:capacity}

An important open question is how sequential retrieval reliability scales with the number of stored memories and the length of the retrieved sequence. In classical Hopfield networks, the synaptic matrix is fixed, whereas in the proposed slow-fast model it is dynamically modulated by the saliency variables:
\[
W_{\mathrm{Hopfield}}
=\frac{1}{N}\sum_{\mu=1}^{P}\xi^\mu(\xi^\mu)^\top,
\qquad
W(|z|)
=\frac{1}{N}\sum_{\mu=1}^{P}|z_\mu|\,\xi^\mu(\xi^\mu)^\top.
\]
In the classical setting, all stored memories contribute simultaneously to cross-talk. In contrast, during sequential retrieval, only a small subset of saliency coefficients typically remains significant, while the others decay toward zero. Thus, the nominal memory load $P/N$ need not reflect the number of memories effectively contributing to the synaptic dynamics at a given time.

To investigate this effect, we consider two numerical protocols and measure the probability of successfully completing an entire prescribed retrieval cycle, following the sequential retrieval criterion of~\cite[Eq.~(4)]{TC-HZV-JA:24}. Successful retrieval requires the prescribed sequence of memory transitions, rather than merely escaping from the current memory state. In the \emph{growing-cycle} protocol, a directed cycle of length $L=\operatorname{round}(P/4)$ is embedded among $P$ stored random i.i.d. memories with $\mathbb{P}(\upxi^\upmu_i=\pm 1)=1/2$, with $N\in\{256,512\}$. In the \emph{fixed-cycle} protocol, the cycle length is kept at $L=5$ while the memory load $P/N$ is increased substantially, with $N=256$. Both protocols are evaluated with SmoothF and Tanh activations ($\kappa=15$ and $Z_0=8$), using 10 independent memory realizations per configuration.

\begin{figure}[!ht]
    \centering
    \includegraphics[width=0.9\linewidth]{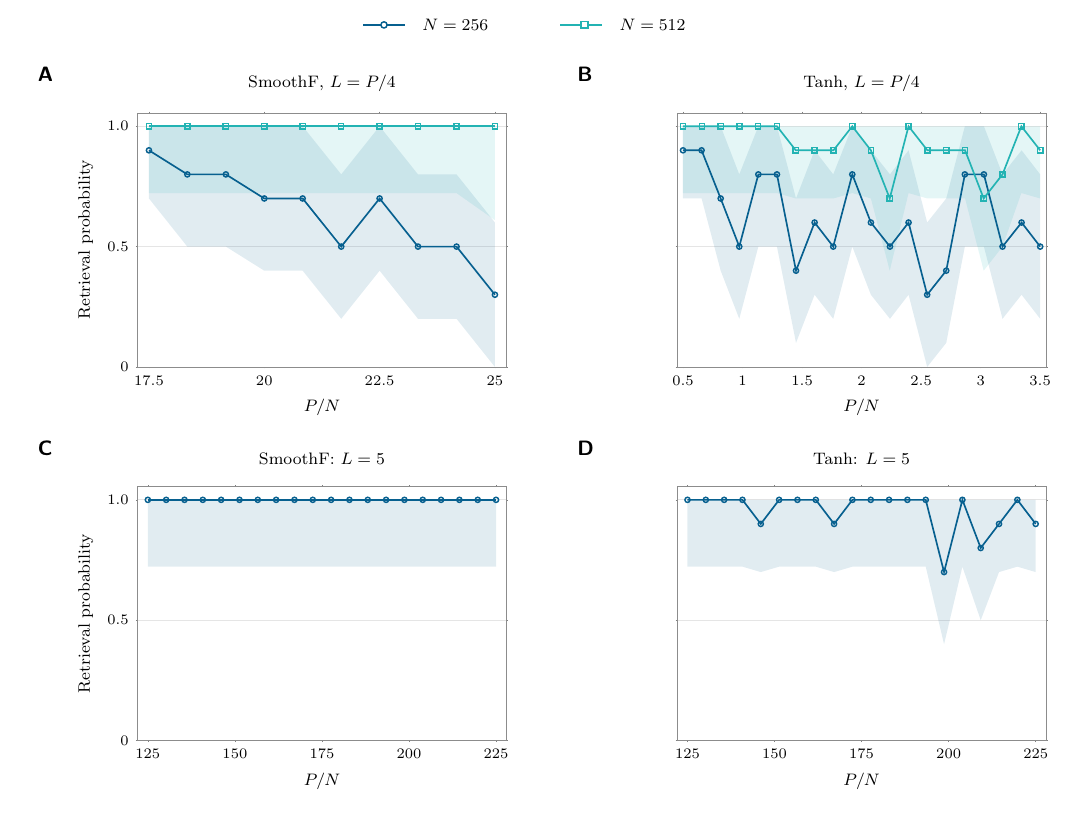}
    \caption{Full-sequence retrieval probability versus memory load $P/N$.
    (A, B) Growing cycles with $L\simeq P/4$ and $N\in\{256,512\}$ for SmoothF and for Tanh, respectively.
    (C, D) Fixed cycles with $L=5$ and $N=256$, for SmoothF and Tanh, respectively.
    Shaded regions indicate 95\% confidence intervals across independent memory realizations.}
    \label{fig:app-capacity-cycles}
\end{figure}

Figure~\ref{fig:app-capacity-cycles} reports the resulting retrieval probabilities. In the growing-cycle setting (panels A,B), SmoothF achieves reliable retrieval $P/N\in(17.5,25)$ and $N=512$, while retrieval probability constantly decreases for $N=256$ over the studied memory load range. Tanh exhibits greater variability, particularly for smaller network sizes. In contrast, with a fixed cycle (panels C,D), retrieval remains highly reliable even for $P/N\gg 1$, with occasional failures for Tanh. This difference reflects, in part, the increasing number of correct transitions required when $L$ grows with $P$, whereas the additional memories in the fixed-cycle setting remain largely inactive under the saliency dynamics.

These results highlight the distinction between nominal memory load and the number of memories actively involved in sequential retrieval. In particular, reliable retrieval of a fixed cycle at high $P/N$ does not imply that all stored memories can be reliably retrieved. Rather, it suggests that dynamic saliency modulation can mitigate interference from memories outside the active sequence. Moreover, since both memory load and sequence length increase in the growing-cycle protocol, their individual effects cannot be isolated from these experiments. A precise theoretical characterization of sequential retrieval capacity therefore remains an open problem.

\section{Branching memory transitions}
\label{app:branching}

Our theoretical analysis focuses on transition graphs with at most one outgoing edge per memory, which enables a scalar reduction of the slow saliency dynamics and explicit conditions for sequential retrieval. However, this restriction is not inherent to the proposed model, which also accommodates branching graphs with multiple possible successors. In this setting, several successor saliencies are driven simultaneously, and branch selection depends on their competing dynamics.

To investigate this behavior, we consider two directed three-memory cycles sharing a common source memory, with equal outgoing transition weights. We use the SmoothF activation, set $N=256$, $P=5$, $\kappa=15$, and generate 50 independent sets of random i.i.d.~$\pm 1$ memories. In the deterministic setting, the selected successor is the one with the larger absolute overlap with the source memory in all 45 realizations with unequal overlaps. We then introduce noise of amplitude $10^{-1}$ per integration step in the slow dynamics and perform 20 trials per memory realization. The successor with the larger absolute overlap remains the more frequently selected branch in 36 out of 43 realizations with unequal overlaps.

These results indicate that branch selection can be influenced by the geometry of the stored memories, even when outgoing transition weights are identical. In particular, differences in memory overlaps can break the symmetry between competing successors and favor one transition over another. This preference persists in most noisy trials, although stochastic perturbations can alter the selected branch. Thus, branching introduces a competition between successor memories that depends not only on the transition graph, but also on memory correlations and perturbations. A general characterization of this competition would require extending the scalar slow reduction to account for the simultaneous evolution of multiple successor saliencies.

\section{A motivating example for input-controlled transitions: Associative Memory Finite State Machine (AM-FSM)}\label{app:input-ex}

Finite state machines (FSMs) are canonical models of sequential computation, in which the evolution of the system is jointly determined by current inputs and a compact internal representation of past states. At any time, an FSM is in one state drawn from a finite set, and incoming inputs trigger transitions according to a predefined state-dependent transition structure. In this sense, FSMs implement a form of implicit memory: past information is compressed into discrete internal states that condition future behavior. The state-dependent interpretation of inputs makes FSMs natural abstractions for a wide range of engineered systems, including elevators, network protocols, and traffic lights with pedestrian signals. In all such cases, effective control requires integrating current observations with the system’s internal configuration, thereby generating context-appropriate actions over time. More broadly, decision-making processes exhibit a similar structure: actions depend not only on present stimuli, but also on the internal state shaped by prior events. This observation motivates viewing sequential decision-making as the evolution of an FSM.

As an illustrative example, consider a train driver operating a metro line in the Tokyo subway system. Under normal conditions, the driver repeatedly traverses a fixed route (the "{\color{BrickRed} red-black}" line), corresponding to a stable sequence of internal states. Due to an unexpected strike, however, the driver receives instructions to deviate from the habitual route and instead pass through Shinjuku station (on the "{\color{OliveGreen} green-black}" line) and subsequently Tokyo station (on the "{\color{MidnightBlue} blue-black}" line). Only two transfer hubs permit switching from the red-black line to the other lines. At each hub, the driver must choose between alternative turns, using prior knowledge of the network and the current location to guide the decision. Although the map is outdated and uncertainty remains, the driver assigns high confidence to a left turn at each hub, while maintaining a lower probability for the alternative right turn.

This scenario exemplifies sequential decision-making under uncertainty, where each choice depends on both external cues and an internally maintained representation of the route taken so far. In the figure below, we instantiate the driver’s behavior using a two-timescale IDP Hopfield model, illustrating how stable internal states and slower adaptation processes jointly support FSM-like dynamics. In particular, we instantiate an input-driven temporal network over the transition graph, where each outgoing edge from a memory $\upxi^\upnu$ towards a set of memories $\{\upxi^\upsigma\}_\upsigma$ is suitably weighted by an external input $u$. The magnitude of the input components along each transition edge will ultimately determine the sequential memory path that is traversed by the two-timescale IDP Hopfield model. The exemplified control over sequential memory transitions provides solid insp
\begin{figure}[!h]
    \centering
    \includegraphics[width=1\linewidth]{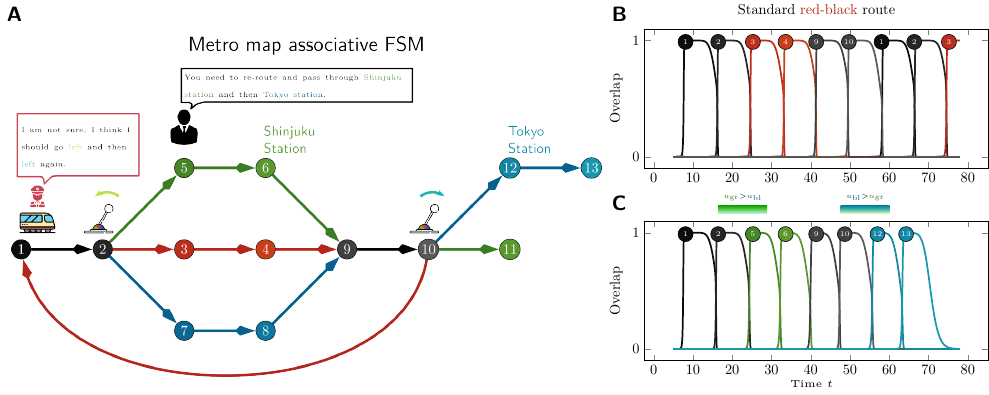}
    \caption{\emph{Finite state machine dynamics with uncertain inputs realized by a two-timescale IDP Hopfield model.}
(A) Schematic of a simplified Tokyo subway graph comprising three metro lines connected by two hub stations. The metro driver (red marker) must adapt the route in response to an external rerouting request, while facing uncertainty at each hub regarding the correct branch to take. 
(B) Sequential retrieval of stable memory states corresponding to stations along the habitual red-black line, illustrating deterministic FSM dynamics under standard operating conditions. 
(C) Input- and state-dependent sequential retrieval during rerouting: external inputs bias the transition dynamics at each hub toward the green-black line (via Shinjuku station) and subsequently the blue-black line (via Tokyo station), while preserving residual uncertainty across alternative transitions.}
    \label{fig: AM-FSM}
\end{figure}

\section{Experimental setup}\label{app:exp}

The code of the experimental campaign is available at \url{https://github.com/sim1bet/Memory-Flows-NeurIPS2026}.

We integrate the fast neural state and slow saliency dynamics by explicit Euler steps on the dimensionless fast clock. For $\tau_x/\tau_z=0.01$, an integration step $\Delta s$ corresponds to slow time $\Delta\theta=0.01\Delta s$; the numerical step and horizon are specified separately for each experiment in \texttt{configs/paper/} and fast noise with amplitude $\upsigma=10^{-6}$. 

The experiments are implemented in the accompanying repository through separate configuration-driven simulation and plotting scripts (\texttt{experiments/}, \texttt{plotting/}); the master command \texttt{run\_all.sh} reproduces the configured workflows. For all experiments, successful numerical retrieval requires a persistent target alignment of at least $0.90$, separated from the runner-up by at least $0.10$, over a $0.05$ slow-time hold window in the capacity tests. Uncertainty is computed over independent pattern sets, not over multiple source trajectories as if they were independent data sets.

All numerical campaigns for the camera-ready figures were executed on the \emph{Leonardo} high-performance computing system, on the AI Factory project EHPC-AIF-2026PG02-1086. The repository includes resolved experimental configurations, seed policies, CSV outputs and figure-generation scripts; numerical parallelism and worker counts are configurable through the command-line interface. Figure generation reads saved experimental data and does not rerun the underlying dynamics.

\end{document}